\documentclass[twocolumn]{aastex62}
\usepackage{hyperref}
\usepackage{ulem}
\pdfoutput=1 
\usepackage{amsmath,amstext}
\usepackage[T1]{fontenc}
\usepackage{apjfonts} 
\usepackage[figure,figure*]{hypcap}
\usepackage[graphicx]{realboxes}

\shortauthors{Boyden et al. 2017} 

\begin{document}


\title{Assessing the Impact of Astrochemistry on Molecular Cloud Turbulence Statistics}

\author{Ryan D. Boyden}
\affil{Department of Astronomy and Steward Observatory, The University of Arizona, 933 North Cherry Avenue, Tucson, AZ 85721 USA}
\email{Email: rboyden@email.arizona.edu}
\affil{Department of Astronomy, University of Massachusetts, Amherst, 710 N Pleasant St, Amherst, MA 01003 USA}
\author{Stella S. R. Offner}
\affil{Department of Astronomy, University of Massachusetts, Amherst, 710 N Pleasant St, Amherst, MA 01003 USA}
\affil{Department of Astronomy, The University of Texas at Austin, 2515 Speedway, Stop C1400, Austin, TX 78712 USA}
\author{Eric W. Koch}
\author{Erik W. Rosolowsky}
\affil{Department of Physics, University of Alberta,  Edmonton, T6G 2E1, Canada}

\begin{abstract}
We analyze hydrodynamic simulations of turbulent, star-forming molecular clouds that are post-processed with the photo-dissociation region astrochemistry code {\sc 3d-pdr}. We investigate the sensitivity of 15 commonly applied turbulence statistics to post-processing assumptions, namely variations in gas temperature, abundance and external radiation field.  We produce synthetic $^{12}$CO(1-0) and CI($^{3}$P$_{1}$-$^{3}$P$_{0}$) observations and examine how 
the variations influence the resulting emission distributions. To characterize differences between the datasets, we perform statistical measurements, identify diagnostics sensitive to our chemistry parameters, and quantify the statistic responses by using a variety of distance metrics. We find that multiple turbulent statistics are sensitive not only to the chemical complexity but also to the strength of the background radiation field. The statistics with meaningful responses include principal component analysis, spatial power spectrum and bicoherence. A few of the statistics, such as the velocity coordinate spectrum, are primarily sensitive to the type of tracer being utilized, while others, like the 
$\Delta$-variance, strongly respond to the background radiation field. Collectively, these findings indicate that more realistic chemistry impacts the responses of turbulent statistics and is necessary for accurate statistical comparisons between models and observed molecular clouds.
\end{abstract}

\nopagebreak[4]
\section{Introduction}

The formation and evolution of molecular clouds is shaped by the interplay between turbulence, gravity and magnetic fields. Observational and numerical studies suggest that stellar feedback\textemdash a process in which forming stars inject energy into their local environment\textemdash also influences cloud energetics across a broad range of scales \citep{frank14}.
The combination of these processes introduces {\it complex, hierarchical} velocity and density structure. 

A variety of 2D and 3D statistics have been developed to analyze spectral data cubes, quantify cloud structure and compare regions. The best statistics are able to extract underlying physical properties like the velocity power spectrum \citep{brunt13} and respond predictably to changes in bulk system properties, such as the turbulent Mach number, influence of gravity or magnetic field strength \citep{burkhart09,correia14,burkhart15,Koch17}.
Numerical simulations play an important role in testing and calibrating statistical metrics. In turn, comparisons between observations and simulations provide a means to evaluate the accuracy of initial conditions and theoretical models.

Missing information, line-of-sight projection and non-linear radiative transfer effects complicate the reconstruction of physical quantities directly from cloud emission \citep[e.g.,][]{beaumont13}. Consequently, effective comparison of observations and numerical simulations requires ``synthetic observations,'' whereby simulations are post-processed to predict the emission assuming the model is a real object, observed by a specific instrument at some point in the sky \citep{goodman11}. 
Over the last decade the necessity of ``apples-to-apples'' comparisons has been widely accepted and increasingly pursued. A burgeoning number of studies have produced synthetic observations and compared with observational data, efforts that have been facilitated by the public availability of a variety of radiative transfer codes \citep{robitaille11,dullemond12,brinch10}.

Meanwhile, a large body of work has been devoted to developing and evaluating the efficacy of different statistics in across a range of physical regimes \citep[e.g., see][and citations therein]{Koch17}. However, few of these studies concern cross-comparisons between different statistics. Recent work by \citet{yeremi14} and \citet{Koch17} has made significant headway towards rigorous comparisons of common metrics with the objective of unifying approaches within a single framework. Such efforts give increasing confidence in the robustness of the ``yardstick'' by which data comparisons are made. 

However, evaluating the success of these efforts and relevance of a given statistic is not always straight-forward. Agreement between a particular dataset and a given synthetic observation does not automatically confer correctness on the underlying model and its specific properties: a range of physical conditions may be degenerate for a given statistic, or worse, may be ill-suited to certain regions of parameter space \citep{burkhart13b,Boyden16,Koch17}. 
Finally, details of the post-processing pipeline differ between studies, and a variety of short-cuts are often adopted. Exactly how these assumptions impact the simulation-observation agreement has not been well-characterized. In the present study, we consider this last factor on the production of synthetic observations.

Ideally, the post-processing pipeline between the model and synthetic observation accurately accounts for dust or molecular abundances, radiative transfer and instrumental limitations, including noise, resolution and completeness. Some studies perform full radiative transfer, while others assume local thermodynamic equilibrium or adopt the local velocity gradient (LVG) approach. Hydrodynamic codes that calculate chemistry in-situ eliminate one factor in the pipeline, albeit by adopting simplified chemical networks \citep{glover11,bertram14,bertram15b}.
Alternatively, stand-alone astrochemistry codes are used to compute the three-dimensional temperature and abundance distribution in post-processing, allowing full-networks and more complex chemistry to be considered \citep{Bisbas12,Offner14b}. 
Most simply, many studies continue to adopt simple functions for abundance and temperature distributions \citep{krumholz07b,offner09a,burkhart13a,Boyden16}.

The final piece of the pipeline involves instrumental modeling. 
This can take a variety of forms, including channel/band-pass modeling \citep{Offner12,koepferl16a}, beam convolution \citep{Offner08,Offner12} and synthesizing noise characteristics \citep{koepferl16a}.
In comparisons with interferometry data, where emission is absent on specific scales, accounting for the telescope configuration and sensitivity is particularly crucial \citep[e.g.,][]{Offner12b,mairs14,dunham16}. 

In this study, we assess the impact of the chemistry pipeline step on synthetic observations and the resulting statistical characteristics.
This paper builds on several previous papers that harnessed synthetic CO observations of numerical simulations to explore astrostatistics. 
\citet{yeremi14} developed and tested a framework for comparing statistical metrics based on experimental design. \citet{Koch17} developed a statistical analysis package, {\sc turbustat}\footnote{\textcolor{red}{\url{http://turbustat.readthedocs.io}}}, and used it in combination with experimental design to assess the responsiveness of a variety of common metrics to physical parameters like Mach number, driving scale, plasma beta parameter and virial parameter. Finally,
\citet[][henceforth B16]{Boyden16} applied {\sc turbustat} to simulations of stellar winds interacting with molecular clouds and investigated the responsiveness of each statistic to stellar feedback. However, all these studies assumed constant molecular abundance. Understanding how gas chemistry affects the statistical responses is a natural continuation of this prior work: this study provides a more realistic test of simulation-observation comparisons and the efficacy of statistical diagnostics. 

In this paper, we focus on three major comparisons: astrochemistry versus no astrochemistry, carbon monoxide versus neutral carbon emission as a gas tracer, and standard UV external radiation field versus an amplified UV field. We analyze the outputs of 
hydrodynamic simulations of turbulent star-forming molecular clouds performed by \cite{Offner13}. \cite{Offner14b} post-processed these simulations to obtain gas temperatures and chemical abundances assuming an external irradiation. 
Here, we produce synthetic observations representing the nearby Perseus molecular cloud and apply the {\sc turbustat} packaged developed by \citet[][henceforth K17]{Koch17}. 
Together, the statistics characterize the data and distill complex emission information into more manageable one-dimensional or two-dimensional forms. We then examine the response of the statistics with and without abundance variation, with and without temperature variation and different cloud irradiation for both CO and atomic Carbon emission. \S\ref{method} discusses our approach in more detail. \S\ref{plots} presents the statistics and discusses the statistical responses to our chemistry parameters. 
\S\ref{distance} compares the distance metrics and quantifies the statistical responses that we describe in \S\ref{plots}. We discuss the results in \S\ref{discuss} and summarize our conclusions in \S\ref{conclusions}. We note that \S\ref{plots} and \S\ref{distance} contain similar content, but they highlight different components of our analysis. If the reader is primarily interested in examining the statistical diagnostics of astrochemistry parameters, we suggest focusing on \S\ref{plots}. For those interested in seeing what the distances look like for our simulation suite, we suggest reading \S\ref{distance}.

\section{Methodology}\label{method}

\subsection{Hydrodynamic Simulations}

We analyze a hydrodynamic simulation of a turbulent, star-forming molecular cloud performed by \cite{Offner13}, which these authors labeled Rm6. We provide a brief summary here and refer the reader to that paper for numerical details. The calculation was performed with the {\sc ORION} adaptive mesh refinement code. It assumes periodic boundary conditions and represents a 600 M$_\odot$ piece of a Milky Way cloud. The domain size of 2 pc and the base grid size is 256$^3$. The calculation has four AMR levels, for a minimum resolution of 100 AU. The gas velocity is initialized using a random field with power on large scales (wavenumbers $k= 1-2$) and a global velocity dispersion of $1.25$ km/s. 
For our model velocity and density distribution, we 
choose a snapshot at one free fall time, during which 17.6\% of the gas has collapsed into stars. 

\subsection{Synthetic Observations}

We adopt chemistry calculations from \cite{Offner14b}, where the 3D photo-dissociation (PDR) code, {\sc 3d-pdr} \citep{Bisbas12}, is applied to the hydrodynamic simulations described above. The network contains H, He, C, O, Mg, S and Fe species and the networks are evolved assuming an isotropic radiation field irradiates the cloud at the domain boundaries. We consider both 1 Draine and 10 Draine fields, where 1 Draine is the standard unit defined by \citet{Draine} and represents the typical local radiation field in the Milky Way.
In these calculations, the elemental abundances, ionization rates and linewidths that are typical for star-forming clouds \citep[see][for further details]{Offner14b}.

Next, we use {\sc radmc-3d}\footnote{\url{http://www.ita.uni-heidelberg.de/~dullemond/software/radmc-3d/}} to compute the $^{12}$CO (1-0) and CI ($^{3}\mathrm{P}_1$-$^{3}\mathrm{P}_0$) emission. Prior observational and numerical studies have shown that these tracers have similar emission distributions and reliably trace H$_2$ abundances in diffuse molecular gas \citep[e.g.,][]{oka01,papadopoulos04,Offner14b, glover15}. 
To solve the equations of radiative statistical equilibrium, we adopt the Large Velocity Gradient approach \citep[e.g.,][]{shetty11}, and we input the tracer abundances and temperatures that were computed with {\sc 3d-pdr} and the densities and velocities from the {\sc orion} simulation. The resulting position-position-velocity (PPV) spectral cubes have a velocity range of 
$\pm 10$ km~s$^{-1}$ and spectral resolution of $\Delta v = 0.078$ km~s$^{-1}$. {\sc radmc-3d} does not consider the freeze-out of CO onto dust grains, which begins to occur around $n_{\rm H_2} \sim 10^4$ cm$^{-3}$, so the chemical abundances in the densest regions are overestimated. Table \ref{simprop} lists the approximate mean optical depth calculated from the {\sc radmc-3d} level populations, gas temperatures and abundances. The optical depth of the emission line along a particular line of sight is given by
\begin{equation}  
\tau = \sum_i \frac{A_{ul}c^3}{4(2\pi)^{3/2}\mathcal{M} c_s  \nu_{ul}^3} n_u \left( \frac{n_l g_u}{n_u g_l}-1 \right) \Delta x_i,
\end{equation}
where $\nu_{ul}$ is the transition frequency, $\mathcal{M}$ is the simulation Mach number, $c_s$ is the sound speed,
$\Delta x_i$ is the cell width, $A$ is the Einstein coefficient, $n$ is the number density, and the subscripts $u$ and $l$ denote the upper and lower levels, respectively \citep{tielens05}. This estimate assumes a Gaussian line shape.
For our synthetic clouds, the emission is typically quite optically thick through the cloud center, optically thin around the edges and marginally optically thick on average. 

Finally, we set the PPV cubes at the distance of the Perseus molecular cloud, $d=$250 pc, and add simulated instrumental noise (${\sigma}_{rms}$ = 0.1 K for $^{12}$CO, and ${\sigma}_{rms}$ = 0.15 K for CI), which is typical for local regions like the Perseus molecular cloud \citep{ridge06}. 
We convert intensity to an antenna temperature by using the Rayleigh-Jeans approximation, and we assume the aperture/antenna efficiencies are unity.
The peaks of bright lines  
in our models range from $\sim 20-50~\mathrm{K}$, so we calculate peak signal-noise of $\sim 100-500$ in our PPV cubes.

\begin{deluxetable}{lcccc}
\tablecaption{Model Properties \tablenotemark{a} \label{simprop}}
\tablehead{ \colhead{Model} &  
   \colhead{Line} &
   \colhead{UV Field} &
   \colhead{Chemistry} &
  \colhead{$\bar \tau$}	
   }
\startdata
CO1DAT  & $^{12}$CO(1-0) &  1 Draine & AT & 1.5 \\
CO1DT  & $^{12}$CO(1-0) &  1 Draine & T  & 0.2 \\
CO1D  & $^{12}$CO(1-0) &  1 Draine & Full {\sc 3d-pdr} & 3.6 \\
CO10D  & $^{12}$CO(1-0) & 10 Draine & Full {\sc 3d-pdr}& 8.5 \\
C1DAT  & CI($^{3}$P$_{1}$-$^{3}$P$_{0}$) &  1 Draine & AT & 12.6\\
C1DT  & CI($^{3}$P$_{1}$-$^{3}$P$_{0}$) &  1 Draine & T  & 4.6\\
C1D  & CI($^{3}$P$_{1}$-$^{3}$P$_{0}$) &  1 Draine & Full {\sc 3d-pdr} & 13.7\\
C10D  & CI($^{3}$P$_{1}$-$^{3}$P$_{0}$) &  10 Draine & Full {\sc 3d-pdr} & 0.3 \\ 
\enddata
\tablenotetext{a}{Model name, emission line, external radiation field, chemical complexity and mean optical line 
depth for each model. Here, 1 Draine (1D) is the typical interstellar radiation field strength as described by \citet{Draine}. A 10D model has a 10 Draine external field. Names containing ``AT'' refer to models that assume uniform abundance and gas temperature. Models with ``T'' only have uniform temperature but the abundances vary assuming an external field as computed by {\sc 3d-pdr}. ``Full {\sc 3d-pdr}'' describes models with local variation in both gas abundances and temperatures as computed by {\sc 3d-pdr}. The mean optical depth is computed by averaging over all lines of sight, where the average excludes sight-lines if the line center optical depth has $\tau < 10^{-2}$. }
\end{deluxetable}

Table \ref{simprop} provides a description of our model suite, which examines different degrees of chemical complexity. 
Four models adopt the temperature and abundances as computed by {\sc3d-pdr} in the radiative transfer step, two models assume uniform abundance and temperature (CO1DAT and C1DAT), and two models have abundance variation but uniform temperature (CO1DT and C1DT). The models with uniform density and temperature have a constant CO abundance of $10^{-4}$ per H$_2$, a constant carbon abundance of $10^{-4}$ per H$_2$ \citep{frerking82}, and an average gas temperature of 22 K, which is the mean gas temperature calculated by {\sc 3d-pdr} 
with a 1 Draine field. 
The models with constant temperature only use the {\sc 3d-pdr} computed  abundances but assume all the gas is 22 K. 

\begin{figure*}[htbp]
	\vspace{-0.1 in}
	\centering
	\includegraphics[width = 1.0 \linewidth]{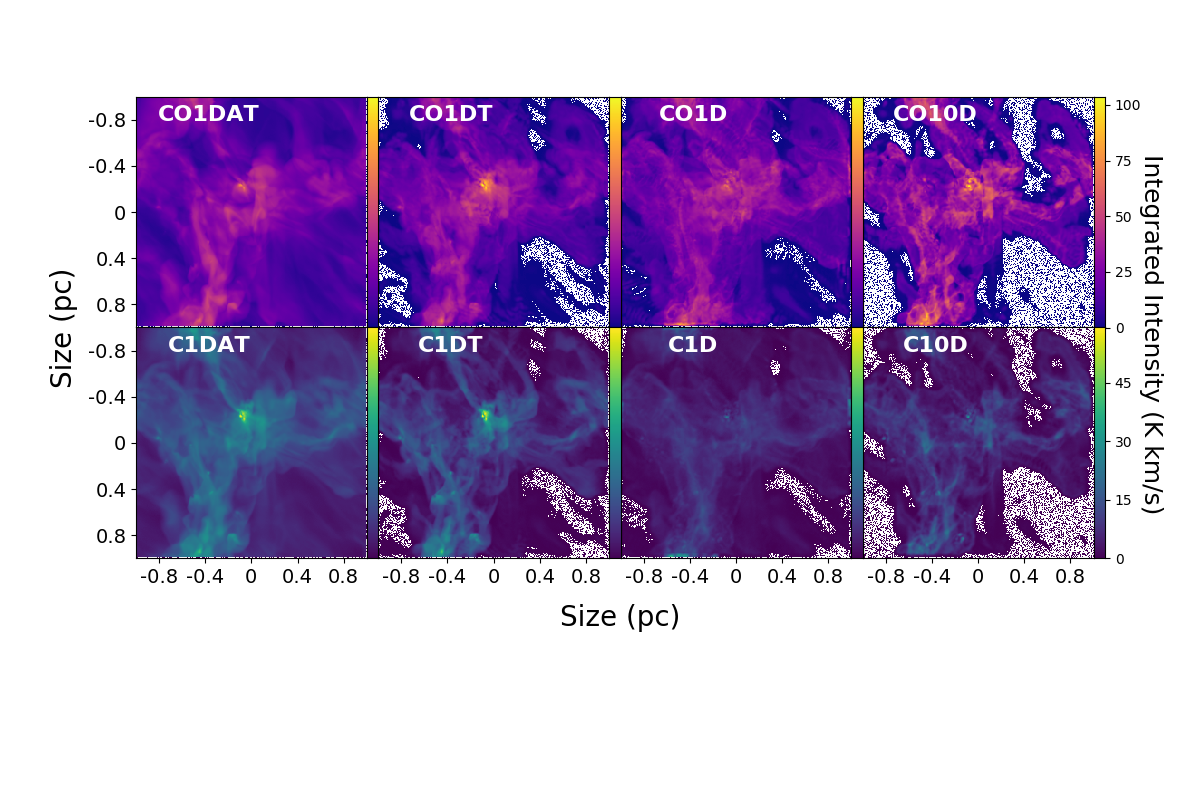}
    \vspace{-1.2 in}
	\caption{Integrated intensity maps for all models. In order to simplify comparisons between models, CO runs (top row) and C runs (bottom row) are plotted with different color-
maps. White pixels correspond to intensity values below the noise threshold, and we omit them from our analysis. 
}
	\label{int_ints}
\end{figure*}

Figure \ref{int_ints} displays the integrated intensity for each model. In general, the CO models are brighter than the C models, and the CO models show more filamentary substructure in the more diffuse cloud regions than the C models. 
Both tracers follow similar trends with the chemistry parameters. Emission is lost in the cloud periphery as a result of photo-dissociation when the gas abundances vary.
When temperature variation is included, less emission comes from the well-shielded gas in the cloud center, which is $\sim 10$ K and colder than the mean temperature, and more comes from warmer gas near the cloud boundary. For C1D and CO1D, the increasing emission due to the temperature variation mostly offsets the reduction in the abundance caused by dissociation. Some gas remains well-shielded and cold even 
in the presence of a 10 Draine field,
although the mean temperature more than doubles to 57~K.
 As the chemical complexity increases, the emission in the central regions becomes more structured. However, the large-scale features of the simulated cloud are retained throughout the post-processing. The importance of these two competing factors depends on the statistical metric of comparison.

\subsection{Statistical Toolkit}

\begin{deluxetable*}{llll}
\tablecaption{ Statistics \tablenotemark{a} \label{stats}}
\tablehead{ \colhead{Family } & \colhead{Name} & \colhead{Comparison Metric}  & \colhead{Key Citations}}
\startdata
 		& Probability Distribution Function (PDF) & Histogram\tablenotemark{b} & \citet{nordlund99} \\
	& PDF Skewness					& Histogram\tablenotemark{b}  & \citet{kowal07,burkhart09} \\
Intensity 	& PDF Kurtosis 					& Histogram\tablenotemark{b} & \citet{kowal07,burkhart09} \\
Statistics  		& Principal Component Analysis (PCA)	& Eigenvalues & \citet{heyer97,brunt02a, brunt02b,brunt13} \\
		 & Spectral Correlation Function (SCF) 	& Surface & \citet{rosolowsky99,padoan99} \\ \hline
  		&  Spatial Power Spectrum (SPS) & Power-law Slope\tablenotemark{b}  & \citet{lazarianp04} \\
     		&  Velocity Channel Analysis (VCA) & Power-law Slope & \citet{lazarianp00,lazarianp04}\\ 
Fourier    & Velocity Coordinate Spectrum (VCS) & Power-law Slope & \citet{lazarianp04,lazarianp06}
\\ 
Statistics 		& Bispectrum & Bicoherence Matrix\tablenotemark{b} & \citet{burkhart09,burkhart10} \\ 
		& $\Delta$-Variance & Spline Fit\tablenotemark{b} & \citet{stutzki98,ossenkopf08a,ossenkopf08b} \\ 
		& Wavelet Transform & Power-law Slope\tablenotemark{b} &  \citet{gill90}\\  \hline
		&  Genus & Spline Fit\tablenotemark{b}  & \citet{gott86,kowal07}\\  
Morphology& Dendrogram Leaves & Power-law Slope & \citet{rosolowsky08,goodman09} \\  
Statistics	 & Dendrogram Feature Number & Histogram & \citet{burkhart13b} \\
\enddata
\label{table2}
\tablenotetext{a}{List of all statistics we calculate, grouped by family. The third column indicates the form of the pseudo-distance metric used to assess the degree of difference between two data cubes. The mathematical definition for each is given in K17. The last column lists seminal papers that have either developed or explored this statistic in detail in the context of molecular clouds.}
\tablenotetext{b}{Statistics that are performed using the 2D integrated emission rather than the full 3D spectral cube.}
\end{deluxetable*}

We use the {\sc turbustat} python package developed by K17 for our statistical analysis.  The package is publicly available and unifies 18 common turbulence statistics under a single framework that enables quantitative comparisons between datasets. It contains algorithms for the following statistics: probability distribution function (PDF), PDF skewness, PDF kurtosis, principal component analysis (PCA), spectral correlation function (SCF), Cramer statistic, spatial power spectrum (SPS), velocity channel analysis (VCA), velocity coordinate spectrum (VCS), modified velocity centroid (MVC), bispectrum, $\Delta$-variance, wavelet transform, genus, dendrogram leaves, dendrogram feature number and Tsallis statistic.
See K17 for the definitions and complete references to the original formulations of these approaches.
 
K17 construct a pseudo-distance metric for each statistic. Pseudo-distance metrics provide a framework by which the sensitivity of different statistics to physical parameters can be quantified. They encapsulate the comparison in one number, which quantifies the difference between two datasets. Common examples of distances include the Euclidean norm, Hellinger norm, weighted difference in slope, and a summed weighted point-by-point difference. We caution that a physical parameter may impact a statistic in multiple ways, but the corresponding distance metric will reflect only certain changes based on its definition, e.g., the slope or the distribution shape. K17 carefully select the distance metrics best-suited for each {\sc turbustat} statistic, and in some cases, like the PDF, different options are possible. {\sc turbustat} also allows two different possibilities for normalizing the statistics -- here we normalize the absolute values of the data before computing the distances to remove variations caused by the different brightnesses of the CO and C emission. 

Table \ref{table2} lists all the statistics that we consider in our analysis. 
Following the results of B16, we exclude MVC and Tsallis statistics. 
Unlike B16, we also exclude the Cramer statistic, as K17 finds it to be sensitive to trivial offsets and noise. Our CO and C cubes have different noise levels, and the CO emission is typically brighter than the C emission. Consequently, we expect the Cramer statistic will likely respond to these effects rather than model parameters.

Table \ref{table2} groups the statistics based on their definitions.  Intensity statistics quantify the emission distributions, Fourier statistics analyze $N$-dimensional power spectra obtained through spatial integration techniques and morphology statistics characterize the structure of the emission (B16). The table also displays the distance metric associated with each statistic and provides relevant citations. Depending on its definition, a statistic is applied to the 3D PPV cube or evaluated using the 2D integrated-intensity map.

K17 examined the stability of the distance metrics for these statistics, in order to ensure that they are sensitive to underlying physics rather than random fluctuations driven by the physical processes. Their approach confirms stability for the formulations, but the statistics could be sensitive to parameters not tested for. 
We refer the reader to that paper for further details.  In this analysis, we consider only the view along the $z$-direction.

\begin{figure}[htbp]
	\centering
	\includegraphics[width = 1.0\columnwidth]{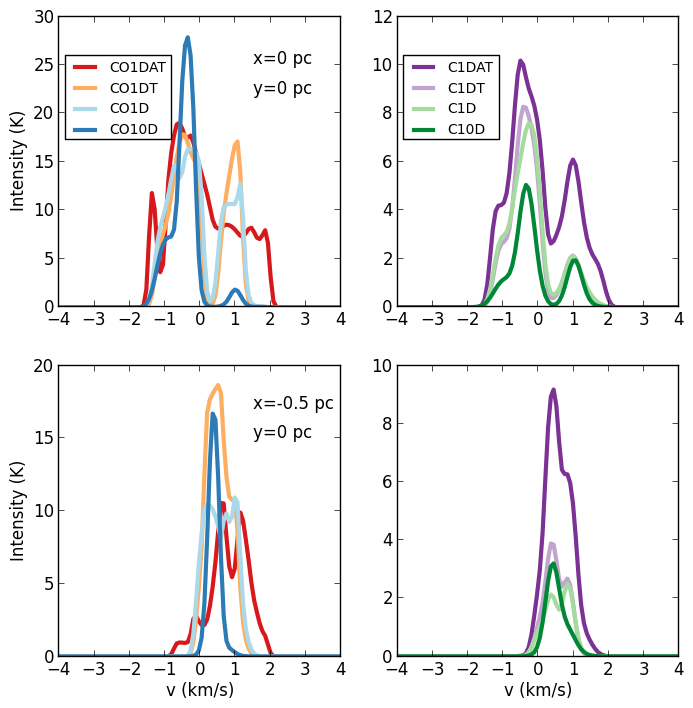}
	\caption{ Intensity as a function of velocity for each model at $(x,y)=(0,0)$~pc, i.e., the middle of the cloud (top panels), and $(x,y)=(-0.5,0)$~pc. The estimated optical depths are [4.8, 0.4, 5.9 and 1.1] (top) and [2.8, 0.6, 5.6, 2.0] (bottom) for CO1DAT, CO1DT, CO1D and CO10D, respectively, and [47.2, 6.0, 18.0 and 0.2] (top) and [24.5, 3.0, 5.2, 0.6] (bottom) for C1DAT, C1DT, C1D and C10D, respectively. 
    } 
	\label{spectra}
\end{figure}

\subsection{Parameterization Scope and Optical Depth}

Normalization of the statistics prior to comparison ensures that differences are driven by variation in line shape and properties rather than total emission. By design, all the analysis is performed on the same density and velocity distributions, allowing us to isolate how choices in post-processing impact the results.

One implicit but fundamental parameter of the study is optical depth. Including variation in abundances and temperatures and increasing the radiation field both change  
the typical optical depth. Table \ref{simprop} indicates the mean line optical depth for each model.
AT models adopt the maximum abundance typical in well-shielded molecular regions. Consequently, applying an external field dissociates and ionizes a significant amount of gas near the domain boundary, thereby reducing the overall abundance. The stronger the field, the less neutral, molecular gas is present. Models including temperature variation have a broader range of excitation, which impacts the level populations. For example, the dense gas is cooler, since it is closer to $\sim$ 10 K than the mean temperature of $\sim 22$ K assumed for the ``T'' models, 
while the lower density gas is warmer. The net result is that the gas is slightly brighter in $^{12}$CO(1-0). 

The changes in abundance, temperature and external field cause the line shape to vary significantly between models at a given location. Figure \ref{spectra} shows the line-of-sight spectrum for each model at two different locations. The figure illustrates that the CI lines at a given point are weaker and have a smaller linewidth, which is consistent with observations. The peak of the line depends strongly on the local gas temperature, with warmer gas producing brighter emission. The abundance significantly influences the linewidth, where dissociation and ionization reduce the amount of gas contributing to the line and lead to fainter line-wings. 

Optical depth can have a large impact on statistical responses. This is especially important to bear in mind for $^{12}$CO, which is commonly optically thick \citep{burkhart13a,beaumont13}. Thus, an ancillary goal of this work is to assess the degree line shape and optical depth impact the statistics we consider.

\section{Results}\label{plots}

In this section, we present the statistical responses for all models, and we identify qualitative differences between them. 
Specifically, we examine how the statistics respond to all three chemistry parameters: the inclusion of chemistry, the use of different molecular tracers, and the impact of a stronger radiation field. 

In the following discussion, models labeled with an ``X'' instead of a CO or C refer to results that are common to both tracers,  i.e., ``X1D'' refers to the results for both models computed with full chemistry and an ordinary UV field.

\subsection{Intensity Statistics}


Here we present the results for the intensity statistics: probability distribution function (PDF), PDF skewness, PDF kurtosis, principal component analysis (PCA), and spectral correlation function (SCF). 

\subsubsection{Probability Distribution Function}

The PDF quantifies the spread of intensity values within the data. Prior work has found correlations between the PDF and Mach number \citep{padoan97}, turbulent forcing \citep{federrath08}, gravity \citep{burkhart15} and feedback (B16).
In this analysis, we calculate the PDF for each 2D integrated-intensity map and normalize the distributions by their respective mean intensity values.

\begin{figure}[htbp]
	\centering
	\includegraphics[width = 1.1\columnwidth]{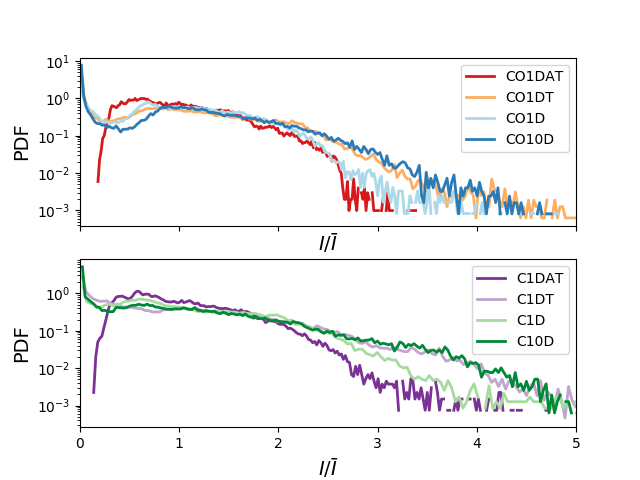}
	\caption{PDFs of the integrated intensity maps. CO models are plotted in the upper panel, while all CI models are plotted on the lower plot.}
	\label{PDF}
\end{figure}

Figure \ref{PDF} shows the computed PDFs, which exhibit similar trends for both CO and CI. 
All models including abundance variation (i.e. {\sc 3d-pdr}) show a sharp increase in the number of pixels at low intensities.  
We note that Figure \ref{int_ints} also displays a low-intensity excess for models including abundance variation: the images of models X1DT, X1D and X10 contain a greater amount of `dark' regions than models X1DAT.
Physically, the low-intensity excess corresponds to the domain boundary where {\sc 3d-pdr} identifies the transition zone between ionized and molecular gas. At the transition zone, photodissociation is maximal. The radiation field dissociated more CO molecules, which leads to a larger quantity of neutral gas. Consequently, we observe less CO emission, so more regions of the cloud appear darker.

CO1DAT and C1DAT, the two models with uniform abundance, have log-normal PDFs with no low-intensity excess. 
Models with an amplified UV field produce the largest asymmetry, since the shape and extent of the cloud boundary depends on how much gas is heated and photo-disassociated by external radiation. 

To quantify differences between the PDFs, we fit each distribution to a commonly-adopted lognormal form \citep{federrath10}. We use a maximum likelihood estimation (MLE) to obtain the best fits to our models, and we compute the PDF distance metric as 
the absolute difference of the fitted distribution widths, $w$, weighted by the quadrature-sum of the fit uncertainties. 
The large spikes at low intensity, although present in our simulations, exist at the noise boundary and are unlikely to be observed. We therefore exclude the tails from our model fitting by imposing minimum intensity cutoffs for the each MLE calculation. 
This allows for effective comparisons of primarily Gaussian behaviors. All measured distribution widths are included in Appendix \ref{appendix}.

We note that, in this paper we choose to use the lognormal width to define the distance metric rather than the Hellinger distance, which was previously adopted by B16
Many observational studies fit column density PDFs to a lognormal model (i.e. \citealt{Lombardi10,Schneider15}), and here we aim to highlight whether common post-processing assumptions alone affect the lognormal fit.

\subsubsection{Skewness and Kurtosis}\label{hstat_sec}

The skewness and kurtosis quantify the shape of the intensity distributions. Defined as the third-order PDF moment, the skewness measures the asymmetry of a distribution. A skewness of zero corresponds to a completely symmetric distribution. 
A positive skewness indicates that a distribution has a longer right side that resembles a tail of positive (high-intensity) values. A negative skewness denotes the opposite: a longer tail of negative (low-intensity) values. The kurtosis is the fourth-order PDF moment, which 
can be used to indicate the outliers of a distribution.
Gaussian distributions have a kurtosis of three. To center the distributions, we subtract three from all values. 
A distributions with a positive kurtosis will contain a greater range of outliers than that of a Gaussian, and so it will appear ``strong-tailed.'' Conversely, a distribution with a negative kurtosis will appear ``weak-tailed'' and have less extreme outliers than that of a Gaussian.

For our comparisons, we measure the distribution of local skewnesses and kurtoses using the integrated intensity maps. We divide the square data grids into smaller circular regions with a radius of 5 pixels, as performed 
in B16 \citep[see also][]{burkhart09}. As Figure \ref{int_ints} shows, several of our integrated-intensity maps contain pixels below the noise threshold. We exclude all noisy pixels from our analysis. So, if a circular region contains at least one noisy pixel, then we exclude that entire circular region from our analysis.

\begin{figure*}[htbp]
	\centering
	\includegraphics[width = 1\linewidth]{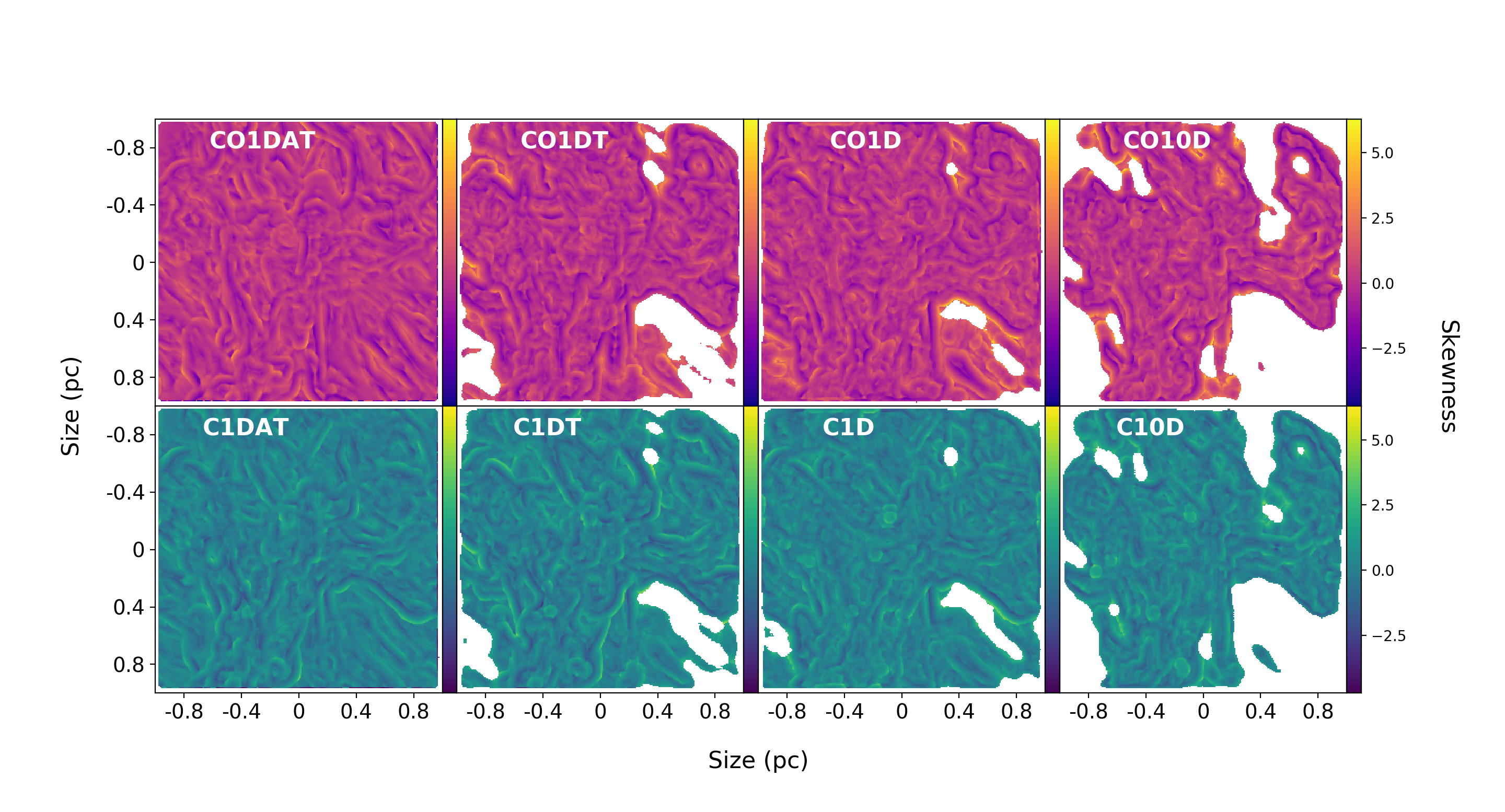}
    \vspace{-0.15 in}
	\caption{ Spatial distributions of the local skewnesses for each integrated-intensity map. At each position, the skewness is computed for all integrated intensities within a radius of 5 pixels. The colorbar denotes the skewness value.}
	\label{skewness_subplots}
\end{figure*}

\begin{figure*}[htbp]
	\centering
	\includegraphics[width = 1\linewidth]{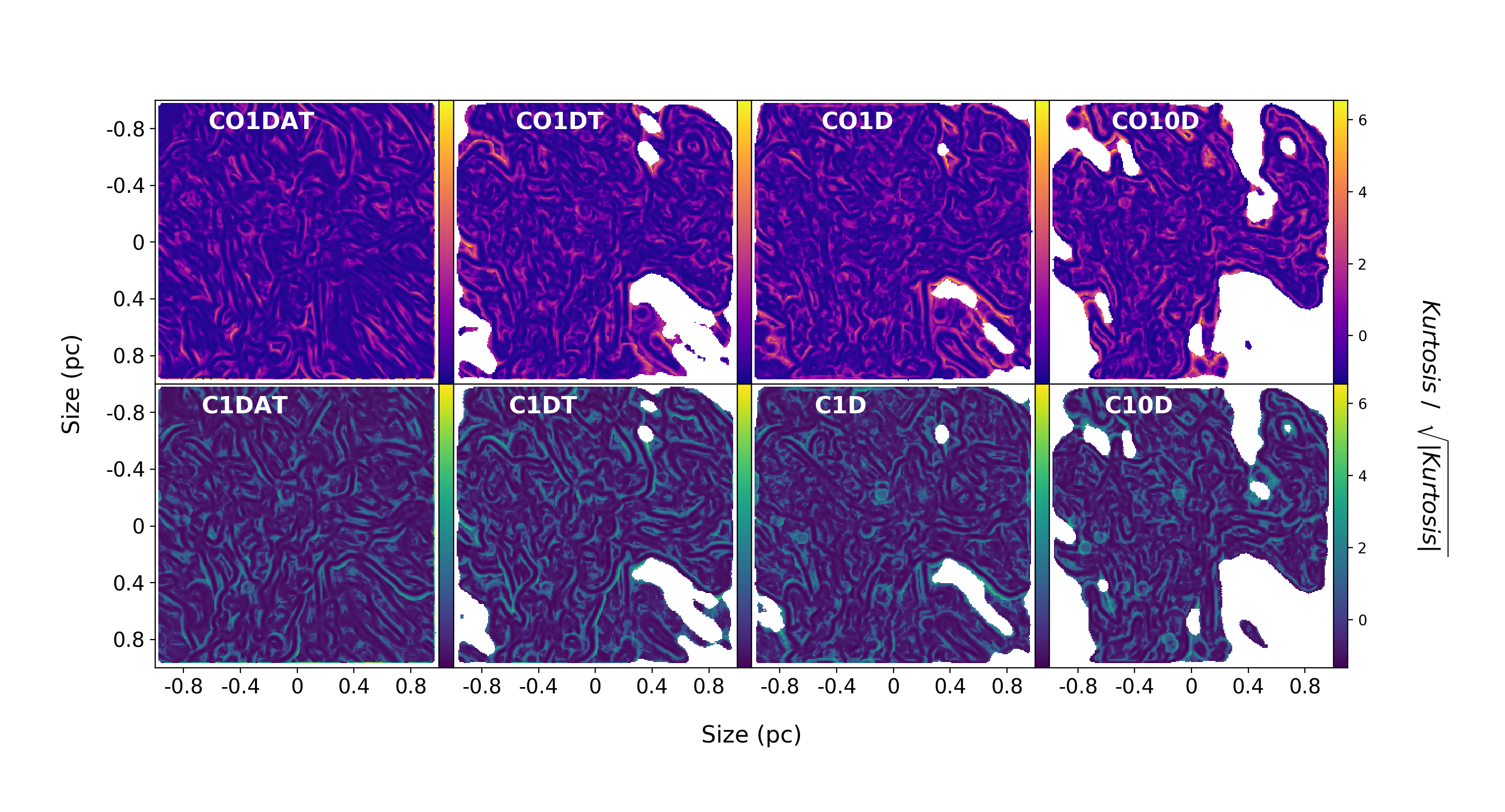}
    \vspace{-0.2 in}
	\caption{Spatial distributions of kurtosis. At each position, the kurtosis is computed for all integrated intensities within a radius of 5 pixels. To emphasize differences in structure, we scale the kurtosis magnitudes by dividing by the square root of the absolute value of the corresponding kurtosis values. 
    }
	\label{kurtosis_subplots2d}
\end{figure*}

\begin{figure}[htbp]
	\centering
	\includegraphics[width = 1\columnwidth]{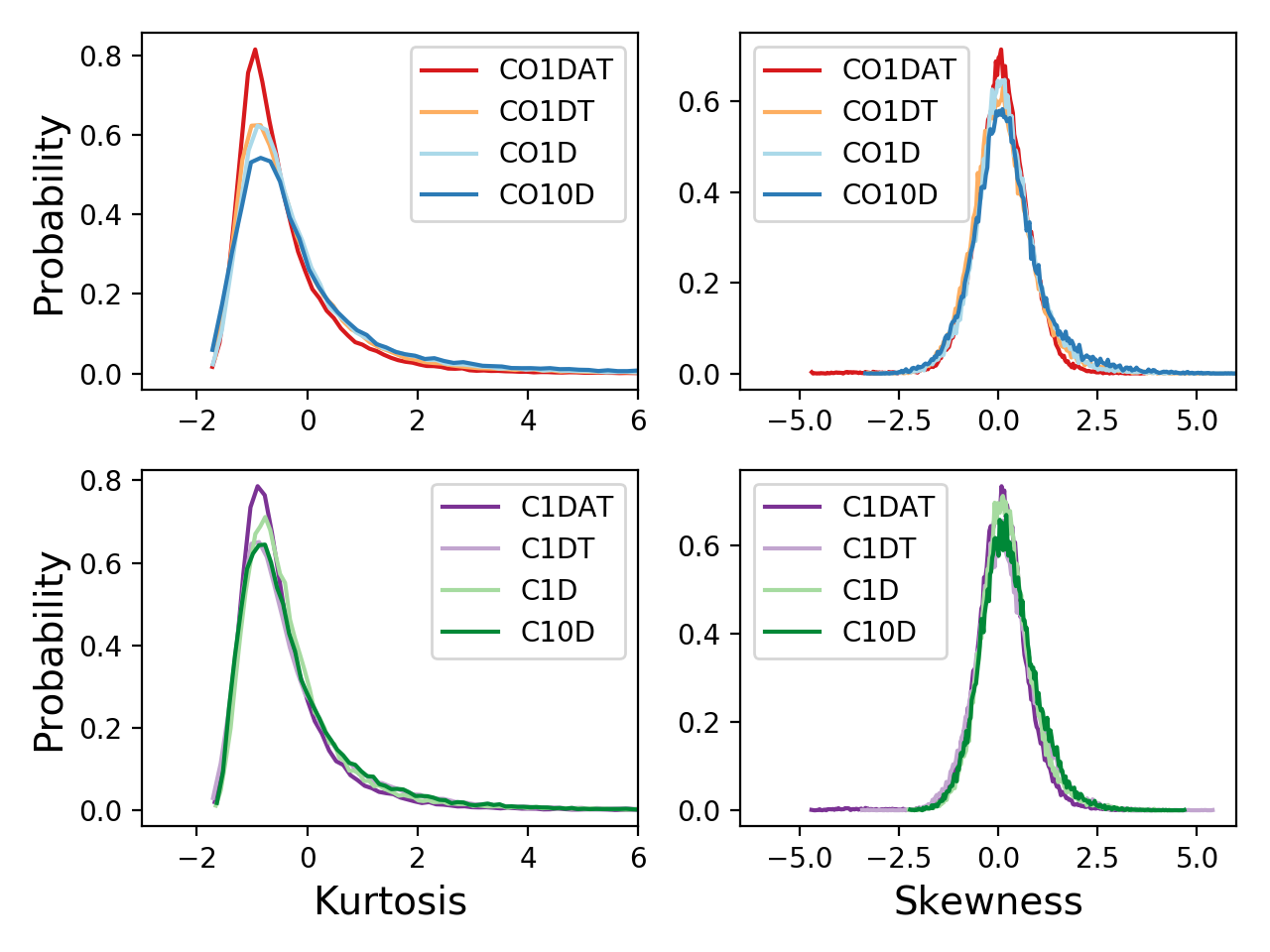}
	\caption{Kurtosis PDFs for the CO models (top-left) and C models (bottom-left); skewness PDFs for the CO models top-right) and C models (bottom-right).}
	\label{highstats}
\end{figure}

In Figures \ref{skewness_subplots} and \ref{kurtosis_subplots2d}, we show the spatial distributions of the skewness and kurtosis, respectively. We find that the cloud structure is visible in both higher-order moments. Namely, the distributions appear tightly correlated with discontinuities (i.e. shocks). For all models, the largest positive skewnesses/kurtoses are often found in filamentary-like structures. 

Figure \ref{highstats} shows the binned skewness and kurtosis values. The skewness PDFs all peak at zero, suggesting that, for a majority of small regions within an integrated-intensity map, the `localized' 
intensity distribution is typically 
symmetric, in part because the lines are typically a symmetric Gaussian. 
The skewness PDFs themselves resemble Gaussians and fall off at smaller or larger skewnesses. 
Models that include abundance variation yield a broader range of positive skewness. We also note slight variations in the distributions peaks, though altogether, the binned skewnesses show weak responses to our simulation parameters.

The kurtosis PDFs behave similarly for all models. They peak at $\sim -1$ and have a tail towards positive kurtosis. 
This trend indicates that the majority of local intensity distributions have extreme outliers, i.e. they are ``strong-tailed.''
As chemistry and a stronger UV field are introduced, the kurtosis PDF peaks flatten, and their tails become more significant. 

A Hellinger norm is used as the distance metric for the skewness and kurtosis statistics. 

\subsubsection{Principal Component Analysis}\label{pca_section}

Principal component analysis (PCA) identifies how a dataset decomposes along a linearly uncorrelated orthogonal basis. 
We use PCA to assess the scale-dependence of turbulent velocity fluctuations of molecular clouds, as described in \citet{brunt13}. 
For our analysis, we calculate a 2D covariance matrix between the velocity channels of the data cubes and determine the corresponding eigenbasis by maximizing the variance along each eigenvector.
The eigenvalues represent the amount of variance projected onto the corresponding eigenvectors, which are denoted as the ``principal components'' of the PPV cubes. 

\begin{figure*}[htbp]
	\vspace{-0.05 in}
	\centering
	\includegraphics[width = 1\linewidth]{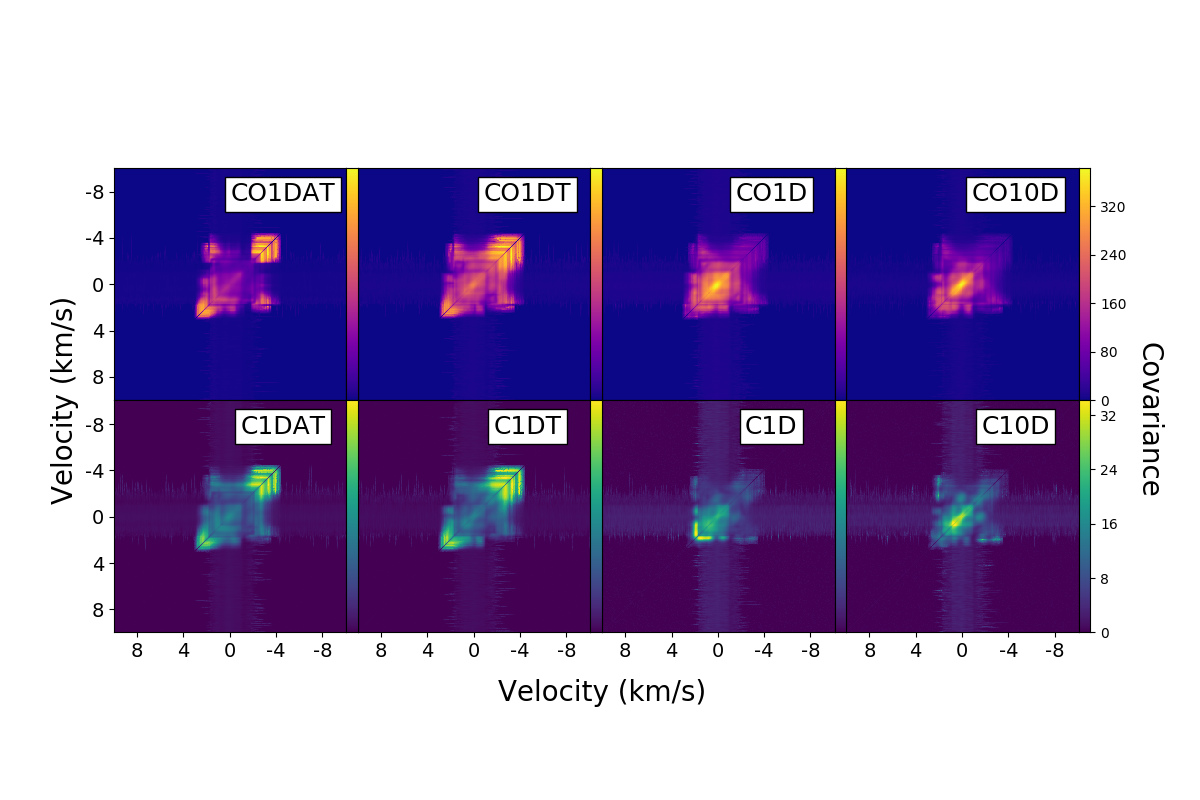}
    \vspace{-0.6 in}
	\caption{Covariance matrices of the velocity channels for all models. The horizontal and vertical axes denote the velocity channels that are used to calculate the total covariance over all spatial positions. The colorbar indicates the covariance magnitude.} 
	\label{pca_surfaces}
\end{figure*}

Figure \ref{pca_surfaces} shows the covariance matrices for each model. We find that all models exhibit covariance across the same range of velocity channels, $|v|<$ 4 
km~s$^{-1}$. 
However, the magnitudes of the covariances strongly depend on whether a model includes temperature variations or not. 
Models X1DAT and X1DT show the strongest covariance at velocities $|v|\sim$ 3-4 
km~s$^{-1}$. 
Having inspected the velocity channels, we find that the emission at $|v|>$ 3 km~s$^{-1}$ is particularly compact:  only a few pixels significantly contribute to the emission, while the rest of the pixels are noisy. At smaller velocity channels, the emission is more spatially distributed.  Velocities $|v|>$ 3 km~s$^{-1}$ correspond to gravitational infall
at the locations of the most massive sink particles in the simulation. 
The covariance peaks for models X1DAT and X1DT trace these infall velocities, which are locally correlated.

We note that the strongest covariances of X1DAT and X1DT are close to but off of the diagonal. Diagonal entries represent the variance of one velocity channel. Since the majority of pixels in the velocity channels $|v|>$ 3 km~s$^{-1}$ are noise, the variances of those channels are small. However, the covariance of two adjacent velocity slices, i.e. $|v| \sim$ 3 km~s$^{-1}$, can still be larger than the variance of one channel. This depends on the relative differences in the emission structure of a dataset. For our data, the compact emission signatures of gravitational infall are, in fact, able to create strong covariances but weak variances at the associated velocity channels.

Models X1D and X10D show the strongest covariances at velocities $|v| < $ 2 
km~s$^{-1}$ as opposed to $|v|\sim$ 3-4 km~s$^{-1}$, and these maxima are along the diagonal.  The emission is more spatially distributed at lower velocity channels, so the variance of one channel can be significant, as opposed to that of high-velocity channels. Furthermore, when gas temperatures vary (in addition to abundance variation), the covariance decreases across the 
velocity channels containing infall ($|v| \sim$  3 km~s$^{-1}$).
This occurs because temperature variation creates more uniform spectra that have Gaussian line shapes, and so the covariance matrix is maximized along the peak of the Gaussian, i.e. the diagonal. 
This trend arises for both gas tracers, though the increased radiation field amplifies the signals of smaller turbulent gas velocities because all of the cold gas is now
spatially co-located at the cloud center.

We have also investigated whether or not the changes in covariance magnitude 
depend 
upon our implementation of PCA. A common 
step in the PCA computation involves subtracting the data by the mean before calculating the covariance matrix. Our implementation, following \cite{brunt13}, omits mean subtraction, since the mean values contribute primarily to the first eigenvalue. For all models, we compared the covariance matrices both including and excluding mean subtraction, and we find no significant differences in their behaviors. Thus, the trends in covariance are indeed properties of our data, which depend on our model parameters.

\begin{figure}[htbp!]
	\centering
	\includegraphics[width = 1\columnwidth]{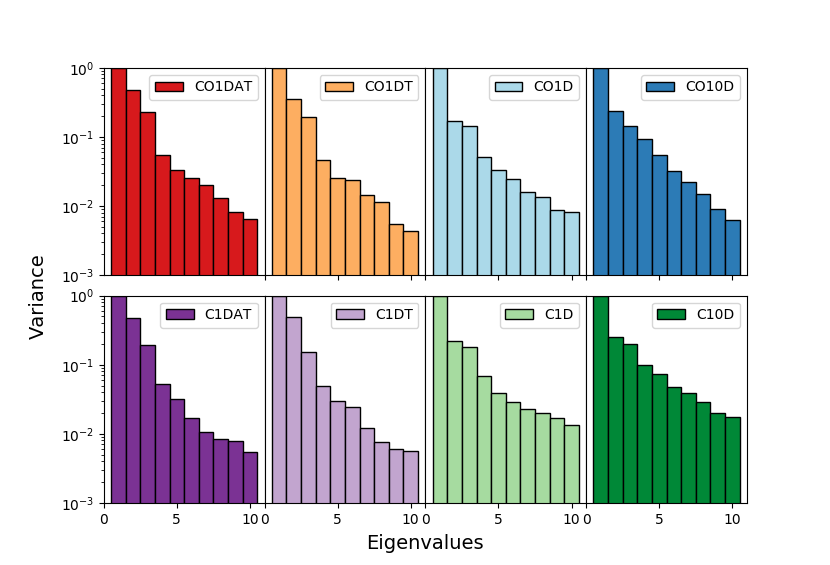}
	\caption{The first 
10 eigenvalues of the covariance matrices of the models normalized to the first eigenvalue. Each magnitude represents the relative variance along the corresponding eigenvector or ``principal component.''} 
	\label{pca_eigs}
\end{figure}

We display 10 
eigenvalues of all covariance matrices in Figure \ref{pca_eigs}. The eigenvalues for CI decline more slowly than those for CO.
We also find the most noticeable differences between the second and third eigenvalues: they are smaller for X1D and X10D and describe less of the variance. This indicates that some larger velocity-structures are lost with the full incorporation of chemistry. When temperature variation is included, C1D and C10D exhibit a significant increase in the variance accounted for by the higher-order principal components. This suggests the emission becomes more structured on smaller scales. In contrast, the eigenvalues are largely insensitive to abundance variations.

The PCA distance metric is defined in {\sc turbustat} as the normalized sum of the difference between the eigenvalues. The first eigenvalue of each covariance matrix correlates with the mean intensity of the corresponding model, since we neglect the mean subtraction step in our PCA implementation. Thus, in order to standardize the data prior to comparisons, we omit the first eigenvalue in our computation of the distance metric. Moreover, higher order eigenvalues mainly reflect the noise in our data, so we choose to compute distances with only 10 lower-order components (excluding the first eigenvalue). Thus, the distances should be especially responsive to the differences between the second and third eigenvalues.

\subsubsection{Spectral Correlation Function}

The spectral correlation function (SCF) was first proposed by \cite{rosolowsky99} and is a measure of similarity between neighboring spectra in a data cube. For our analysis, we follow \cite{yeremi14} and define the SCF as the normalized root-mean-square (rms) difference between two spectra as a function of their projected separation in the sky. 
Under this definition, the values of the SCF are bound between 0 and 1, where a value of 0 indicates no correlation and a value of 1 indicates perfect correlation. 

\begin{figure}
	\centering
	\includegraphics[width = 1\columnwidth]{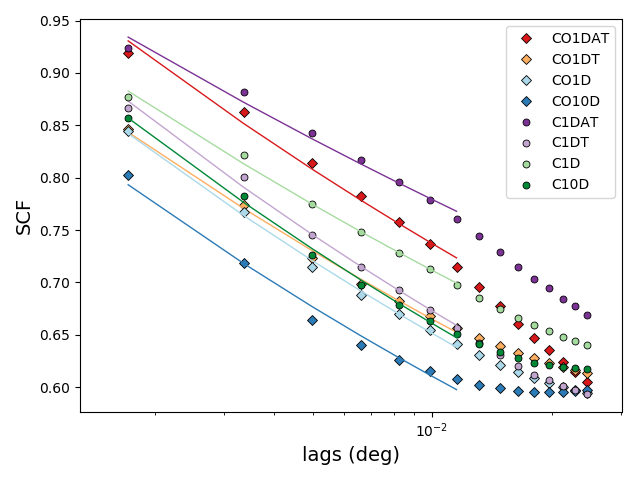}
	\caption{The SCF spectra for all models. These represent the azimuthal average of the SCF surface. The solid lines are power-law fits to the spectra. In our analysis, we restrict the fit interval to [10$^{-3}$, 10$^{-2}$] degrees.
}
	\label{scf_spectra}
\end{figure}

We show the results of the SCF analysis in Figure 
\ref{scf_spectra}, 
which displays the SCF surfaces as a power law. To obtain these, we radially average the SCF surfaces over each projected separation magnitude, which by convention is denoted as the ``lag.'' Because the SCF is a normalized quantity, scale offsets are significant. As we systematically introduce abundance variation, temperature variation, and a stronger radiation field, the SCF decreases in offset  for CO models.
This indicates that both photodissociation and a stronger radiation field reduce spatial correlations in simulated CO emission. 

C emission behaves similar to CO emission with the exception of C1D, which yields a larger SCF values relative to C1DT.  As shown in Figure \ref{int_ints}, C1D has a smoother intensity map and less structure than C1DT does. 
Figure \ref{int_ints} also suggests that there is less of a difference between the emission structure of CO1DT and CO1D.
Physically, temperature variation yields higher temperatures at the domain boundary, and because C has a higher excitation temperature than CO, C emission is strongly associated with the cloud border than CO emission. 
Consequently, the C emission map is smoother, the normalized rms difference between pixels is smaller, and hence its SCF increases over all lags.

Beyond 10$^{-2}$ degrees, the spectra or models with chemistry appear to flatten. This is likely due to missing emission at larger lags (the domain boundary), where photodissociation removes CO and neutral C. At larger scales, models X10D are flatter than models X1DT and X1D,
since the higher photodissociation removes additional CO and C at the cloud edges.

To quantify behaviors away from the cloud boundary, namely, at smaller spatial offsets, we fit the SCF spectra to power laws where we restrict the fit range to [10$^{-3}$, 10$^{-2}$] degrees and report the slopes and uncertainties in Appendix \ref{appendix}. We use an ordinary least squares model to fit the SCF spectra in log-log space.
All of the derived slopes are between $0.1$ and $0.15$, and for both tracers, the slopes increase as we introduce each model parameter. C1D is, again, an exception to the trend, which is likely due to the large slope of C1DT. The overall range of slopes is also less than but comparable to the SCF slope of a synthetic observation of Perseus that was performed by \cite{gaches15}. \cite{gaches15} use the same simulations as we do, but they use the SCF definition given by \cite{padoan03} instead of that of \cite{rosolowsky99}. They also convolve the simulation data with a beam appropriate for FCRAO observations, which we do not. These differences likely explain why \cite{gaches15} report a higher SCF slope of $0.2$. 

The SCF distance metric calculates a normalized pixel-by-pixel difference between two surfaces. Based on Figure \ref{scf_spectra}, models X1DAT and CO10D 
show the most significant differences based on the value of the SCF.  

\subsection{Fourier Statistics}

This subsection presents the results for the Fourier statistics: the spatial power spectrum (SPS), velocity channel analysis (VCA), velocity coordinate spectrum (VCS), bispectrum, $\Delta$-variance, and wavelet transform.

\subsubsection{Spatial Power Spectrum}\label{sps_section}

The spatial power spectrum (SPS) describes the contribution of different spatial scales to the total power or energy. A large body of prior work has characterized the SPS in numerical simulations \citep[e.g.,][]{maclow04,federrath10} and observations \citep[e.g.,][]{burkhart13b} by analyzing the density, velocity, energy or emission.
Here, we compute the SPS by taking the Fourier transform of the integrated intensity maps, calculate the 2D power spectrum, 
and then radially average over spatial frequency to obtain a one-dimensional power spectrum. 

\begin{figure}[htbp!]
	\centering
	\includegraphics[width = 1\columnwidth]{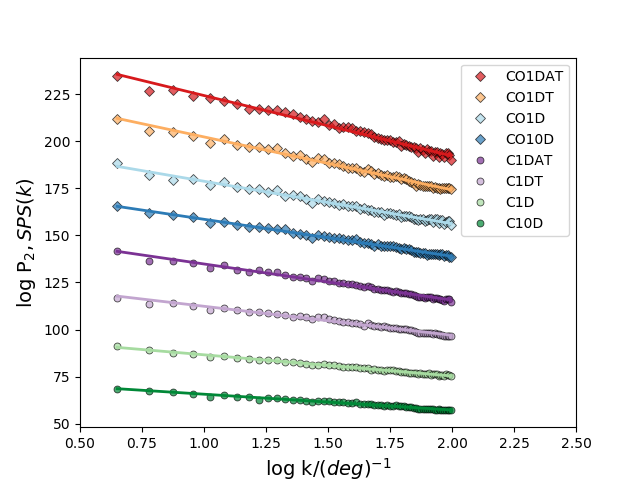}
	\caption{SPS as a function of spatial frequency for all models. The $x$ axis is normalized to units of deg$^{-1}$ and is dimensionless. All the SPS have similar power values, but we offset the SPS in the vertical direction for clarity. The solid lines correspond to power-law fits of each spectrum. 
}
	\label{sps}
\end{figure}

Figure \ref{sps} shows the SPS for all models and the associated power-law fits, which are carried out over the inertial range. For our models, this corresponds to a limit of $k < 2$  $(\mathrm{deg})^{-1}$. The power laws are derived from an ordinary least squares model, and we report the fitted slopes and uncertainties 
in Appendix \ref{appendix}. 
The SPS curves actually have similar energy across all scales; in figure \ref{sps}, we manually offset them  to clearly show differences in the fits. 
Most of the power-law slopes are $\sim -3$, which is consistent with the results of a turbulent optically-thick gas, as described in \citet{lazarianp04} and \citet{burkhart13b}. 
However, models with temperature variation yield shallower slopes ($\sim -2.8$ for X1D and X10D) than those computed with a constant gas temperature ($\sim -3.1$ for X1DAT and X1DT). 
This suggests that the temperature variation has an impact beyond simply changing the optical depth as found by \citet{burkhart13b}. Abundance variation also yields shallower slopes, though this impact is not as significant as that of temperature variation.

In K17, the SPS distance metric is calculated as a t-statistic of the difference in slope. Consequently, the distances between models are likely correlated with temperature variation.

\subsubsection{VCA and VCS}\label{vca_s_section}

Velocity channel analysis (VCA) and velocity coordinate spectrum (VCS) are techniques that isolate how fluctuations in velocity contribute to differences between spectral cubes. 
\citet{lazarianp00} introduced VCA to quantify the impact of turbulent fluctuations on emission lines in turbulent molecular clouds, and \citet{lazarianp04} extended the framework to consider optically thick lines. \citet{lazarianp06} introduced VCS. 

VCA and VCS are derived using similar integration techniques. First, we compute the three-dimensional power spectrum from the Fourier transform of the PPV cube. To obtain the VCA, we integrate the power spectrum over the velocity channels and then radially average over the resulting two-dimensional surface. This results in a one-dimensional spectrum as a function of spatial frequency, which we fit to a power law using an ordinary least squares method. To obtain the VCS, we instead average the three-dimensional power spectrum over the two-dimensional spatial frequencies. The resulting one-dimensional spectrum is a function of velocity-frequency. As described in K17, larger velocity-frequencies, or smaller spectral scales, generally represent density-dominated emission, and smaller velocity-frequencies, or larger spectral scales, generally represent emission from velocity-dominated turbulence. The scales where each component dominates depends on the underlying indices of the three-dimensional velocity and density fields \citep{lazarianp06}. 
To quantify the impact of these two regimes, we follow K17 and use a segmented piece-wise regression model \citep{Muggeo03} to identify a break point and fit two distinct power laws. 

\begin{figure}[htbp!]
	\centering
	\includegraphics[width = 1\columnwidth]{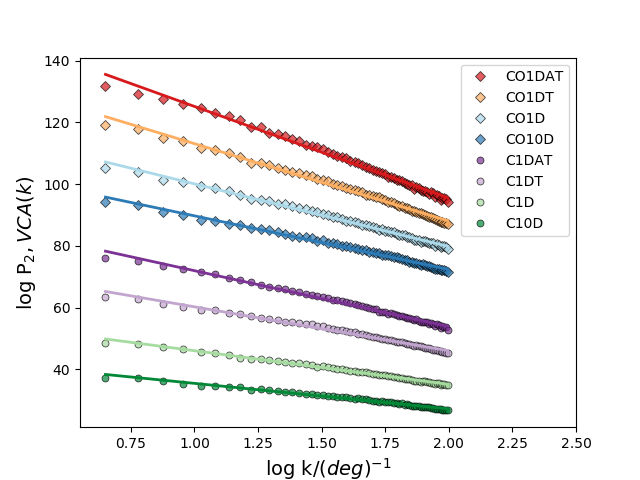}
	\caption{VCA power spectra for all models as a function of spatial frequency. The $x$ axis is normalized to units of deg$^{-1}$ and is dimensionless. The y-values are shifted vertically for clarity. Solid lines are the power-law fits to each spectrum. }
	\label{vca}
\end{figure}

Figure \ref{vca} displays the VCA. 
Similar to the SPS, we perform power-law fits over the inertial driving range, using an upper limit of $k = 2$  $(\mathrm{deg})^{-1}$.
The power-law fits are reported in Appendix \ref{appendix}.
 VCA and SPS both yield 2-dimensional power spectra as a function of spatial frequency, but we report minor differences in the fit results. This is expected, as the SPS only considers spatial information whereas the VCA also considers spectral information, namely, the thickness of the velocity channels.
The power-law slopes for all models are $\sim -3$, and we also find these values to be sensitive to both abundance and temperature variation. For both gas tracers, the slope 
decreases as we introduce each variation. However, the slope differences between models X1D and X10D are statistically insignificant, indicating that the VCA is insensitive to changes in the interstellar radiation field.

The VCA distance metric, which is also a t-statistic of the differences in slope,
yields relatively small differences between the models since the slopes are all similar.

\begin{figure}[htbp!]
	\centering
	\includegraphics[width = 1\columnwidth]{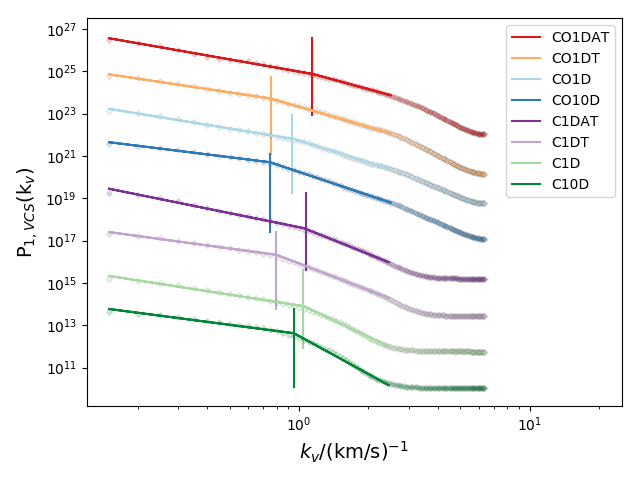}
	\caption{VCS power spectra for all models as a function of velocity-frequency. The $k_v$-axis is normalized to units of (km/s)$^{-1}$ and is dimensionless. The solid lines represent segmented power-law fits. The $y$-axis does not reflect 
actual VCS values: the spectra are 
offset, which allows for better visualization of the fitted data. Colored vertical lines indicate the location of the break point for the corresponding simulated dataset.} 
	\label{vcs}
\end{figure}

Figure \ref{vcs} displays the VCS spectra for the models after  
re-scaling and shifting the VCS to highlight differences in shape. This current scaling highlights a clear difference in the shape of the VCS spectra that \sout{also} depends on the gas tracer. The VCS of the C models are flatter at 
large $k_v$ (small velocity scales). 
However, closer examination shows that these flattened power spectra are noise-dominated, and we exclude these $k_v$ from the segmented power-law fit. We report the results of all segmented power-law fits, including the break point, in  Appendix \ref{appendix}.

The break points of the segmented power-law fits follow similar bulk 
trends for each tracer.
The model break-points shift towards smaller velocity-frequency/larger scales when variations in the temperature and abundance as well as a stronger radiation fields are introduced. 
We note that this trend is not monotonic: models X1D yield larger-valued break points than models X1DAT. Furthermore, while CO10D produces the smallest-valued break point of all CO models, for the C models C1DT does so. The difference between CO and C is likely associated with the weaker response to the radiation field that is exhibited by the C models. As Figure \ref{vcs} shows, the break points for C1D and C10D appear closer together than those of CO1D and CO10D. 
These findings indicate that including chemical modeling and/or a stronger radiation field generally reduces the amount of large-scale emission, but the degree of reduction is likely gas-tracer-dependent.

We also note similar nuances in the behavior of the power-law slopes.
When abundance and temperature variation are included, as well as an amplified radiation field, the slopes in the density-dominated (smaller scales/larger $k_v$) regime get steeper. The net increase in slope from X1DAT to X1D is $-0.41 \pm 0.06$ for CO and $-0.8 \pm 0.1$ for CI. From X1D to X10D the small-scale slopes steepen by $-0.21 \pm 0.6$ for CO and $-0.8 \pm 0.1$ for CI. The large-scale (smaller $k_v$; velocity-dominant regime) slope changes similarly for both gas tracers, but there there is no consistent increase or decrease from X1DAT to X1D. The amplified radiation field causes the slopes to flatten by $-0.42 \pm 0.03$ for CO and $-0.26 \pm 0.06$ for CI.

Because VCS yields two power laws, K17 define the VCS distance metric as the sum of both the t-statistic difference in large-scale scope \textit{and} the t-statistic difference in small-scale slope. The location of the break points is not accounted for in the formulation of the distance metric.

\subsubsection{Bispectrum/Bicoherence}

The bispectrum measures both the magnitude of and phase difference between Fourier signals. Formally, it is defined as the Fourier transform of the three-point autocorrelation function, and it is a unique Fourier statistic because it preserves phase information. The spatial power spectrum, defined as the Fourier transform of the two-point autocorrelation function, does not include information about phase. \cite{burkhart09} introduced the bispectrum as a tool for probing nonlinear wave-wave interactions in a turbulent energy cascade \citep[see also][]{burkhart16}. 
They showed that MHD wave modes produce stronger turbulence correlations than that of purely hydrodynamical perturbations. 

To obtain the bispectrum, we compute the Fourier transform of the integrated intensities in order to construct the three-point correlation function. 
For our analysis, we use the bispectrum to calculate the bicoherence, a real-valued, single-valued, normalized summary that also describes phase information. It is defined as a scalar between 0 and 1, where the value of the scalar gives the degree of correlation between two Fourier signals, i.e. $\vec{k_1}$ and $\vec{k_2}$. A value of 0 indicates completely random phase coupling and a value of 1 indicates complete phase coupling. Moreover, if $\vec{k_1} \neq \vec{k_2}$ 
and also show significant phase coupling, this could suggest a correlation between signals of different phases and/or spatial scales. In the case of molecular clouds, this would indicate the degree to which a hierarchical turbulent cascade is coupled, i.e., the degree to which signals at different spatial frequencies are similar.

\begin{figure}[htbp!]
	\centering
	\includegraphics[width = 1\columnwidth]{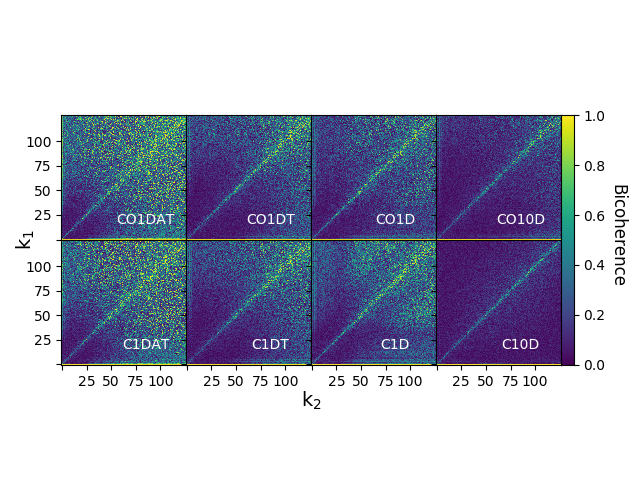}
    \vspace{-0.5in}
	\caption{Bicoherence matrices for all models. Here, the bicoherence is computed over sets of 
randomly sampled spatial frequencies up to $\sim$ half of the image size. 
The colorbar indicates the bicoherence magnitude for each pair of spatial frequency magnitudes, denoted as $k_1$ and $k_2$. A bicoherence of 0 indicates random phase coupling, while a bicoherence of 1 indicates complete phase correlation.}
	\label{bicoh}
\end{figure}

We calculate the bicoherence over sets of 
spatial frequencies, $k_1$ and $k_2$, where each set corresponds to one spatial frequency sampled over 500 random phases. In our analysis, we set the upper spatial frequency limit to $\sim$ half of the image size (127 pixels), and we plot the results as two-dimensional arrays, as shown in Figure \ref{bicoh}. The rows and columns of each subplot display the sampled spatial frequency magnitudes, $k_1$ and $k_2$.  
Each pixel denotes the corresponding bicoherence. We note that, although the sets $k_1$ and $k_2$ are sampled over the same range of magnitudes, they are likely sampled at different phases.

The bicoherence follows the same general behavior, as described in \cite{burkhart09}, for both CO and C emission. All models produce a clear signal across the diagonal, i.e. $k_1 = k_2$, which is expected: this line describes the bicoherence of two identical spatial frequencies, which will always show spatial correlation. Off the diagonal, i.e. $k_1 \neq k_2$, the models show various isocontours, which decrease in bicoherence magnitude as the corresponding spatial-frequency magnitudes decrease (i.e. going from top-right to bottom-left).
The spatial frequencies at which the isocontours sharply decline to zero correspond to the spatial scales at which phase coupling is lost.

Figure \ref{bicoh} shows that abundance variation, temperature variation, and an amplified radiation field all impact the degree to which CO and C models lose phase coupling.  .
Models X1DAT yield the largest amount of correlation:
many spatial-frequency pairs give bicoherence values $>0.4$, and the corresponding isocontours decrease to 0 only at small spatial frequencies (large spatial scales), $k \sim 50$. These matrices resemble the bicoherence values computed in B16, even though these simulations lack magnetic fields and
represent smaller sizes and velocity scales 
than those of B16 and \cite{burkhart09}.
As temperature and abundance variations are included the amount of correlation decreases, since the isocontours decline to bicoherences $<0.2$ at spatial frequencies larger than those of models X1DAT, i.e. $k \sim 75$. This effect is further augmented  
in the transition to an amplified interstellar radiation field: CO10D shows weak isocontours even at the largest spatial-frequencies, $k \sim 100$, and C10D only shows strong correlation along the diagonal. Physically, photodissociation reduces the correlation between emission signals of differing magnitude and phase, especially in the presence of stronger UV radiation. We suggest that more realistic cloud chemistry, as shown here, obscures the phase coupling that arises from other physical cloud parameters, such as the expansion of wind shells (B16) or the propagation of MHD waves \citep{burkhart10}.  

The bicoherence distance is the normalized summed pixel-by-pixel difference in the output matrices. Based on the results in Figure \ref{bicoh}, we expect the metric to respond to our model parameters, especially the interstellar radiation field.

\subsubsection{$\Delta$-Variance}\label{delvar_sec}

\begin{figure}[htbp!]
	\centering
	\includegraphics[width = 1\columnwidth]{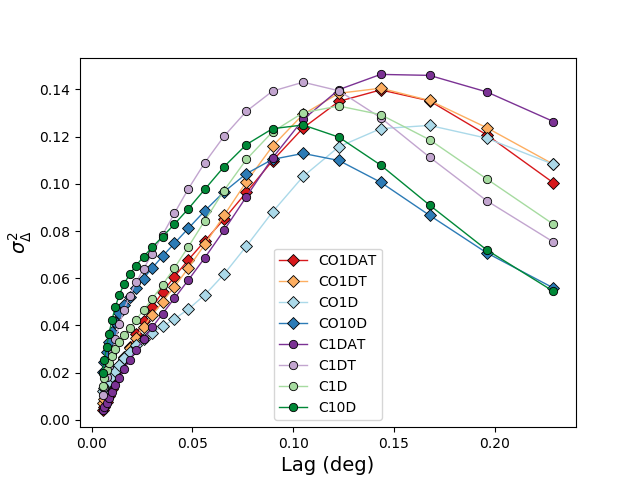}
	\caption{$\Delta$-variance spectra for all models. Here, lag refers to the width of the Mexican-hat wavelet that is used in the calculation. }
	\label{delvar}
\end{figure}

The $\Delta$-variance is a useful tool for characterizing the structure of interstellar turbulence \citep{ossenkopf08a}.
For a given range of spatial scales, the statistic yields filtered maps 
of the emission and then computes a filtered average over the map. 
The filtering assesses how different spatial scales contribute to the total power. We calculate the $\Delta$-variance by using an improved algorithm that was introduced by \citet[][]{ossenkopf08a, ossenkopf08b} and is effective at discriminating between noise and small-scale structure in 2D maps. We generate a series of Mexican-hat wavelets that are approximated as the differences between two Gaussians with a diameter ratio of 1.5, as used in \cite{ossenkopf08a}. The wavelets all vary in width, which is related to spatial scale: different widths filter out different turbulent scales. 
For each model we weight the integrated intensity maps by its inverse variance, convolve it with each Mexican hat wavelet, 
and calculate the $\Delta$-variance. Figure \ref{delvar} displays the $\Delta$-variance spectra for all of the models. Each data point represents a $\Delta$-variance that was computed with a Mexican-hat wavelet of a particular width, denoted by ``lag.'' Since the $\Delta$-variance quantifies the amount of structure present at a particular spatial scale, an entire spectrum reveals the amount of structure that is present at every spatial scale. The spectra shown in Figure \ref{delvar} exhibit a wide range of slopes and behaviors, so we follow B16 and do not fit the spectra to power laws. 

The shapes of all $\Delta$-variance spectra agree with the expected behavior for that of a moment map with variable noise, as described by \cite{ossenkopf08b}.
These spectra are  characterized by a single local maximum, which indicates the spatial scale with the most structure.
For example, \cite{ossenkopf08b} calculated the $\Delta$-variance spectra for rho Ophiuchus and found the maximum to correspond to $0.1$ pc, namely the typical size of cores in that cloud. 
Below the maximum value, the $\Delta$-variance increases steadily with respect to increasing spatial scale. For a moment map, 
these spatial scales are fractal \citep{stutzki98} and are sometimes fit with a power law \citep{ossenkopf08b}. At scales larger than that of the local maximum, the $\Delta$-variance declines as emission structure becomes washed out. 
 
We find various trends in the $\Delta$-variance spectra of our models. 
First, CO1DAT and CO1DT appear nearly identical, suggesting that abundance variation hardly impacts structure in CO emission. This behavior is in contrast with that of models C1DAT and C1DT. In C emission, abundance variation lowers the $\Delta$-variance values at larger lags but increases those at smaller lags. This causes the spectrum to peak at lower spatial lags, which encapsulates the overall change in emission structure. We find that this difference in gas tracers agrees with the structure of the integrated intensity map, as shown in Figure \ref{int_ints}. Abundance variation 
reduces C emission associated with larger spatial scales. Filaments of C emission also become more pronounced with abundance variation, causing the $\Delta$-variance to increase at smaller scales. CO emission is also impacted by abundance variation, but the relative amount of structure at each spatial scale appears unaffected. 

Moreover, including temperature variation decreases the $\Delta$-variance of CO emission at all scales, with the exception of the uppermost spatial lag. We also note that temperature variation has a different effect on C emission. Relative to CO1DAT, CO1D has less structure at larger scales but slightly more structure at small scales. Relative to C1DT, C1D instead shows less structure at smaller scales and more structure at larger scales.
Increasing the radiation field removes a significant amount of large-scale structure in model CO1D and adds a small amount of small-scale structure. 
This behavior is analogous to the effect of abundance variation on C emission: smaller spatial scales begin to dominate the emission structure. A similar trend also happens with C10D but to a lesser degree.

The local maxima of our $\Delta$-variance spectra occur at lags of $\sim 0.1 - 0.16$ degrees, which in our simulations correspond to $0.4-0.7$ pc. We note that 
these spatial scales are larger the $0.1$ pc maximum $\Delta$-variance that was identified by \cite{ossenkopf08b}.
For models X1DT and X1D, the positions of local maximums are tracer-dependent. CI emission exhibits larger $\Delta$-variances at smaller lags, while the corresponding CO emission has larger values at higher lags. This causes the $\Delta$-variance of the CO models to peak at larger lags, which directly correlates with differences in optical depth. On smaller, denser scales, CO is optically thick, so the filtered emission saturates, causing structure to be washed out.

The $\Delta$-variance distance is defined as a Hellinger distance, so the metric will be sensitive to differences in emission structure at all lags.

\subsubsection{Wavelet Transform}\label{wavs}

The wavelet transform is an alternate technique for studying filtered power spectra. \cite{gill90} first introduced the technique on $^{13}$CO data as a means to quantify the fractal structure of the L1551 outflow region. Unlike the auto-correlation function that is used to compute the SPS, VCA, and VCS, the wavelet transform preserves positional information while characterizing emission structure. 
Our calculation of the wavelet transform follows that of \cite{gill90}: we convolve the integrated intensity maps with a Mexican-hat wavelet, and then we compute the transform as the average of the positive values. We note that the wavelet transform complements the $\Delta$-variance, since both statistics quantify a filtered intensity map: the wavelet transform utilizes the average of positive filtered values, while the $\Delta$-variance utilizes the variance of filtered values.

\begin{figure}[htbp!]
	\centering
	\includegraphics[width = 1\columnwidth]{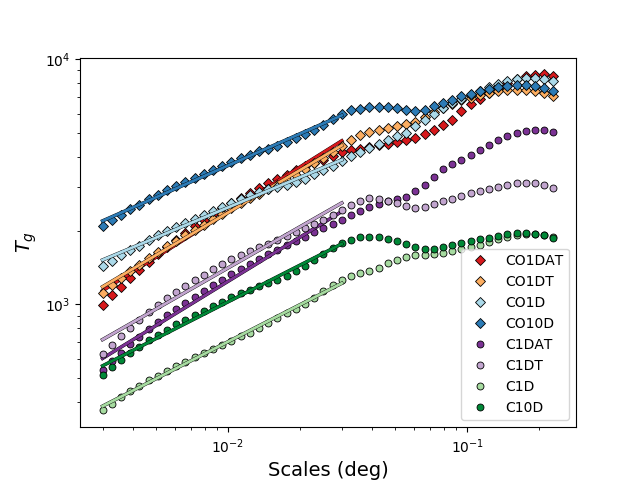}
	\caption{The wavelet transform for all models. The solid lines indicate power-law fits, which all have an upper limit of $0.03$ degrees. These spectra are not rescaled, unlike the SPS, VCA, and VCS.} 
	\label{wavelet}
\end{figure}

Figure \ref{wavelet} shows the wavelet transform of all models. The CO emission  produces higher transform values than those of the CI. 
This result differs from that of the $\Delta$-variance, in part because the wavelet transform reflects the relative brightness of the CO emission. However, like the $\Delta$-variance, structure also decreases on larger scales when the radiation field is increased.

Incorporating abundance and temperature variation impacts the magnitude and shape of the CI wavelet transform but has little effect on the CO wavelet transform.
For the C models, abundance variation reduces the wavelet transform on large scales, while temperature variation significantly reduces the transform across all scales. 
Essentially, this is the signature of there being less emission, which leads to a greater sensitivity in the shape of the curve. 

Both statistics exhibit a similar response to the amplified radiation field: the transform increases at smaller scales, where the emission becomes brighter and a local maximum appears at $\sim 0.03-0.04^{\circ}$. 

We note that the wavelet transform includes information similar to both the $\Delta$-variance and the SPS. Below the local maxima at $\sim 0.04^{\circ}$, the wavelet transform follows a single power law and should be related to the SPS \citep{gill90}. Moreover, the variation in the local maxima roughly matches the variation seen in the $\Delta$-variance peaks in Fig. \ref{delvar} (see also \citealt{stutzki98}).

In this analysis, we use an ordinary least squares method to fit the wavelet transforms with a power law. Here, we restrict the fit range to be below the local maxima, by using at $0.03^{\circ}$ as the upper limit. 
We report the derived slopes and uncertainties in Appendix \ref{appendix}. As expected, the wavelet transform -- at smaller scales -- behaves similarly to the SPS. For both tracers, temperature variation significantly reduces the power-law slopes. Namely, we report slope difference of $\sim 0.2$ between CO1DT and CO1D, and $\sim 0.1$ between C1DT and C1D.

The distance metric is defined as the t-statistic of the difference in the fitted slopes. Because we fit the transforms to smaller scales, we expect the power laws to behave similarly to the SPS.

\subsection{Morphology Statistics}

In this section, we discuss the genus and dendrogram morphology statistics.

\subsubsection{Genus}

The genus statistic characterizes emission structure by measuring the topology of a region. It was first applied to the column density of MHD simulations in \cite{kowal07}. \cite{chepurnov08} extended the technique to observations of the Small Magellanic Cloud. These studies defined the genus as the difference between the number of isolated regions above and below a given density threshold. Under this definition, a high-density region contributes positively to the genus, while a low-density one contributes negatively.

In our analysis, we calculate the genus of the normalized 2D integrated intensity maps. We convolve the maps with a 2D Gaussian kernel of width $1$ pixel to remove the effects of noise and small-scale variation. Next, we compute the genus statistic for 100 equally-spaced intensity thresholds. These values range from 20 percent above the minimum intensity to the maximum intensity in the dataset. 
Our algorithm follows K17, and we define a countable isolated region to be $24$ pixels in size. 
We then fit the resulting distributions to cubic splines.

\begin{figure}[htbp!]
	\centering
	\includegraphics[width = 1\columnwidth]{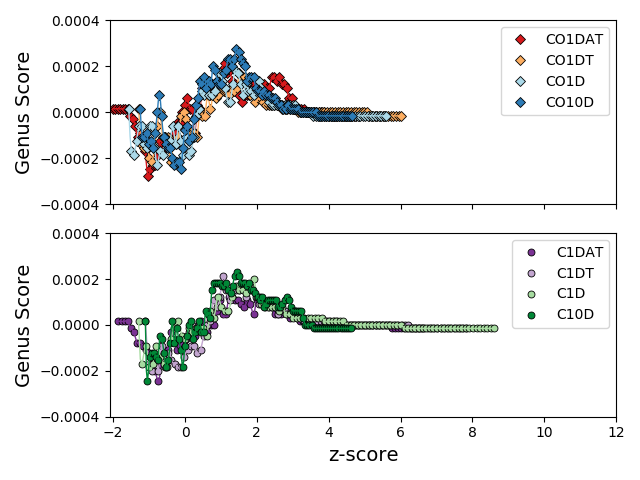}
	\caption{Genus score for all models. The data are normalized such that the $x$ axis indicates the number of standard deviations from the mean intensity. The genus score is normalized by the number of pixels in the integrated intensity map.}
	\label{genus}
\end{figure}

Figure \ref{genus} displays the genus score, normalized by the map area, as a function of standardized intensity, which we denote as ``z-score.'' The genus score analogous to the genus, but it is a function of standardized intensity rather than intrinsic intensity.
Minor fluctuations exist between the models, but we find similar bulk behaviors that agree with those of \cite{chepurnov08}. The curves all contain a local minimum below or near the mean intensity as well as a local maximum at positive z scores $\sim 1.75$. These extrema correspond to the intensity threshold at which there are either the most low-intensity regions (i.e. the minimum) or the most high-intensity regions (i.e the maximum) Further from the extrema, the model topologies become washed out as low- or high-intensity regions begin to coalesce. This causes the genus curves to flatten to zero. 

All model genus curves are negative at the respective mean intensity (z-score of 0), so all of the model topologies are dominated by the low-intensity regions. This appears to be caused by the detailed filamentary structure of our model emission, as shown in Figure \ref{int_ints}. Bright filaments create several isolated low-intensity features, which impact the behavior of the genus curve.

The genus scores of X1D and X10 tend to be greater than the genus scores of X1DAT and X1DT. As Figure \ref{int_ints} suggests, although temperature variation and a stronger UV field does reduce the overall brightness of some regions, they also introduce detailed brightness structure in the form of filaments. This produces a greater amount of isolated high-intensity regions and increases the genus score at most intensity thresholds. We note that the CO models deviate from this behavior. At z-scores of $\sim 0$ and $\sim 4$, as CO1DAT and CO1DT show more bright features than CO1D and CO10D. However, the topology of our models is still dominated by the low-intensity regions rather than the bright filaments, since the genus score is negative at the mean intensity.

Another significant difference between the two gas tracers is the elongated tail in the C models, which indicate that they contain a broader range of normalized intensity values. 
Physically, this also corresponds to bright, compact regions. 
Although the CO models have brighter emission and a greater range of values, 
the standard deviation in their intensity distributions affects the length of the genus tail. Thus, CO emission, though brighter, may yield a narrow genus score if the standard deviation is small. 

The genus distance is defined as the normalized point-by-point difference between two fitted genus curves over their common ranges. 
Thus, the metric only encapsulates differences at standardized intensities $<4$, so it will only highlight some differences between the tails. If the peaks of the genus curves differ, they will dominate the distance instead of the tail.

\subsubsection{Dendrograms}

A dendrogram is a tree diagram that illustrates hierarchical structure in a data set. \cite{rosolowsky08} and \cite{goodman09} demonstrated the utility of dendrograms in characterizing the structure of molecular clouds. 
Following K17, we create dendrograms of the 3D spectral cubes, and  following the approach of \citet{burkhart13b}, we identify the peak intensity value of each PPV cube and catalog all successive local maxima of the intensity distribution. Each peak/local maximum is a leaf, and leaves at similar hierarchies are connected by branches. To ensure each leaf is significant, we set a minimum number of pixels required to define a leaf  \citep{rosolowsky08}. For our analysis, we choose 10 pixels. 
We also define a threshold parameter, ${\delta}_{min}$, which sets the minimum distance between two local maxima \citep{burkhart13b}. Increasing this parameter removes the influence of noise and quantifies how the number of features decreases as smaller-scale structures are removed, namely, by ``pruning'' the tree.

\begin{figure}[htbp!]
	\centering
	\includegraphics[width = 1\columnwidth]{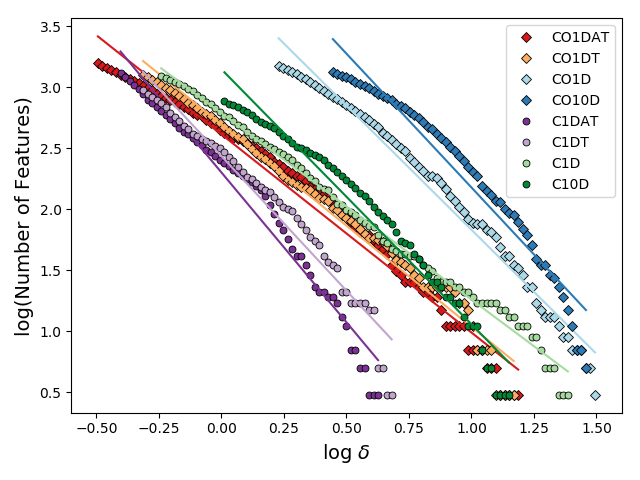}
	\caption{Number of dendrogram features as a function of intensity spacing, $\delta$, for all models. The lines indicate power-law fits. \textcolor{red}{}}
	\label{dendro_num}
\end{figure}

\begin{figure}[htbp!]
	\centering
	\includegraphics[width = 1\columnwidth]{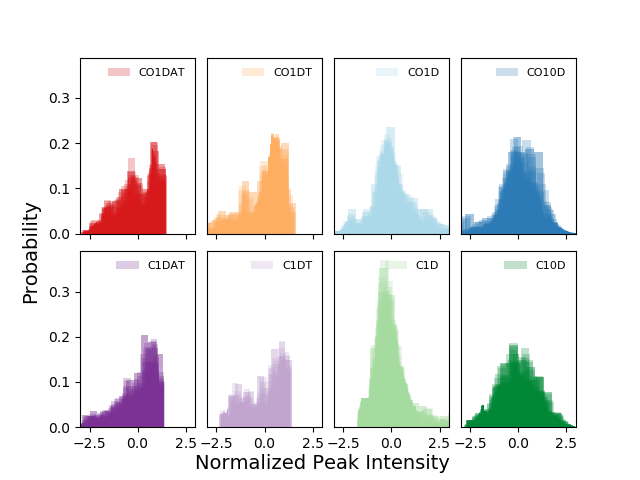}
	\caption{Histograms of the renormalized dendrogram peak leaf intensities. For each value of ${\delta}_{min}$, the resulting transparent histograms are stacked.}
	\label{dendro_hist}
\end{figure}

Following K17, we compute two dendrogram statistics: the number of features statistic and the histograms statistic. The number of features describes how the number of local maxima changes with respect to ${\delta}_{min}$. We compute dendrograms for a series of ${\delta}_{min}$ values that range from $10^{-2.5}$ K to $10^2$ K in 150 logarithmic steps. We record the number of features as a function of ${\delta}_{min}$ and fit a region of the data to a power law, using an ordinary least squares method. The region is defined where ${\delta}_{min}$ exceeds the mean intensity value of a data cube \citep[see][]{burkhart13a}, and it is identified with a sliding window filter function. 
For the same set of ${\delta}_{min}$ values, the histogram statistic computes the distribution of leaf intensities. These intensity values are normalized. 

Figure \ref{dendro_num} shows the number of dendrogram features for all models. The models exhibit similar behaviors: as ${\delta}_{min}$ increases, the number of features declines and eventually falls off sharply. The curves could be fit by two power-laws, but we fit a single one by convention and report the slopes in Appendix \ref{appendix}. Differences between the fitted  slopes vary as a function of dynamic range. Models that span a wider range of ${\delta}_{min}$ values, i.e., CO1DAT, C1DAT, and C1D, are flatter and more like single power-laws than those that span shorter ranges. 

We note that C1D yields the shallowest slope because of its unique behavior at ${\delta}_{min} \sim 10$ K. Here, there appears to be a bump in the power law, where the number of features remains roughly constant over a small range of ${\delta}_{min}$. C1D contains multiple compact bright structures in its central region that are roughly 10 times the mean intensity, and these features are isolated in our computation of the dendrograms. Conversely, CO1D contain the same features, but they are only five times brighter than the mean. 

The dynamic range for each model varies with respect to chemical complexity and gas tracer. Abundance and temperature variation shifts the range to larger values of ${\delta}_{min}$, but abundance variation alone does not significantly affect the statistic. Thus, a broad range of temperatures leads to more excitation, which impacts cloud morphology by  increasing the mean intensity of the models. 

The number of features distance metric is defined as the weighted difference between two slope. Consequently, it primarily responds to differences in the size of the fit range, which correlates with slope. The metric will not respond to the location of the fit range.

Figure \ref{dendro_hist} displays 
the dendrogram histograms. For every value of ${\delta}_{min}$, we plot a transparent histogram. Stacking the histograms conveys the importance of different features in the dendrograms as they are pruned. Altogether,
the CO and C histograms are quite similar, underscoring the similarity in the emission morphology, which has been previously noted \citep[e.g.,][]{papadopoulos04, gaches15}. For uniform temperatures, the CO and CI emission from dense structures becomes optically thick such that the normalized peak intensity limits to a maximum value, truncating the histogram. 
Abundance and temperature variation shifts the histograms from positive skewness to negative skewness, and high-intensity values are less common. An amplified radiation field widens the histogram, indicating an increase in structure at all intensities.

For each ${\delta}_{min}$, the histogram distance metric first calculates a Hellinger difference between two outputs. Then, it adds all of the differences for each threshold, and normalizes the result. Based on the differences in Figure \ref{dendro_hist}, we expect the metric to be sensitive to our model parameters.

\section{Statistical Analysis}\label{distance}

In this section, we quantify all statistic responses to the model parameters. For each statistic, we use the corresponding distance metric to measure the differences between all model pairs 
and gauge sensitivity towards the chemistry parameters, namely, abundance and temperature variation, gas tracer, and interstellar radiation field strength. 
We present the results as color plots as in B16 and K17. 
In the plot, each square indicates the distance between two models as indicated on the horizontal and vertical axes. The color bar denotes the distance, where the range of distance values depends on the statistic.

We organize the models in the same order and systematically introduce more complexity for CO and then for C.
To disentangle the various statistical sensitivities, it is instructive to consider the $8\times 8$ color plots as blocks of four $4\times 4$ matrices. The bottom-left block shows the distances strictly between CO models; the top-right block shows the distances for pairs of C models. The top-left and bottom-right blocks are symmetric and show the distances between models with different gas tracers: a CO model compared to a CI model and vice-versa. 

\subsection{Intensity Statistics}

\begin{figure*}[h!]	
\begin{center}				\includegraphics[width=0.95\columnwidth]{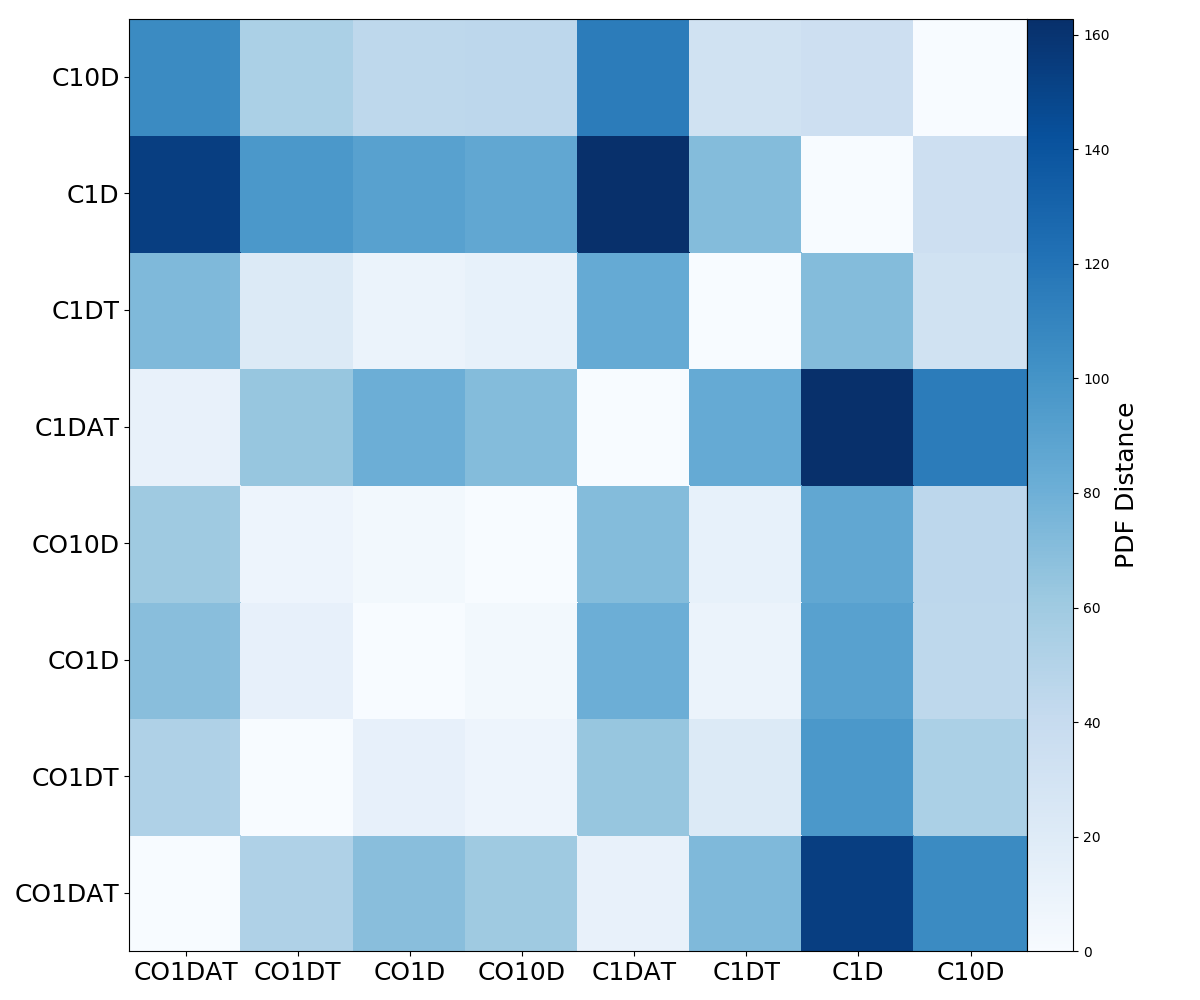} 
\includegraphics[width=0.95\columnwidth]{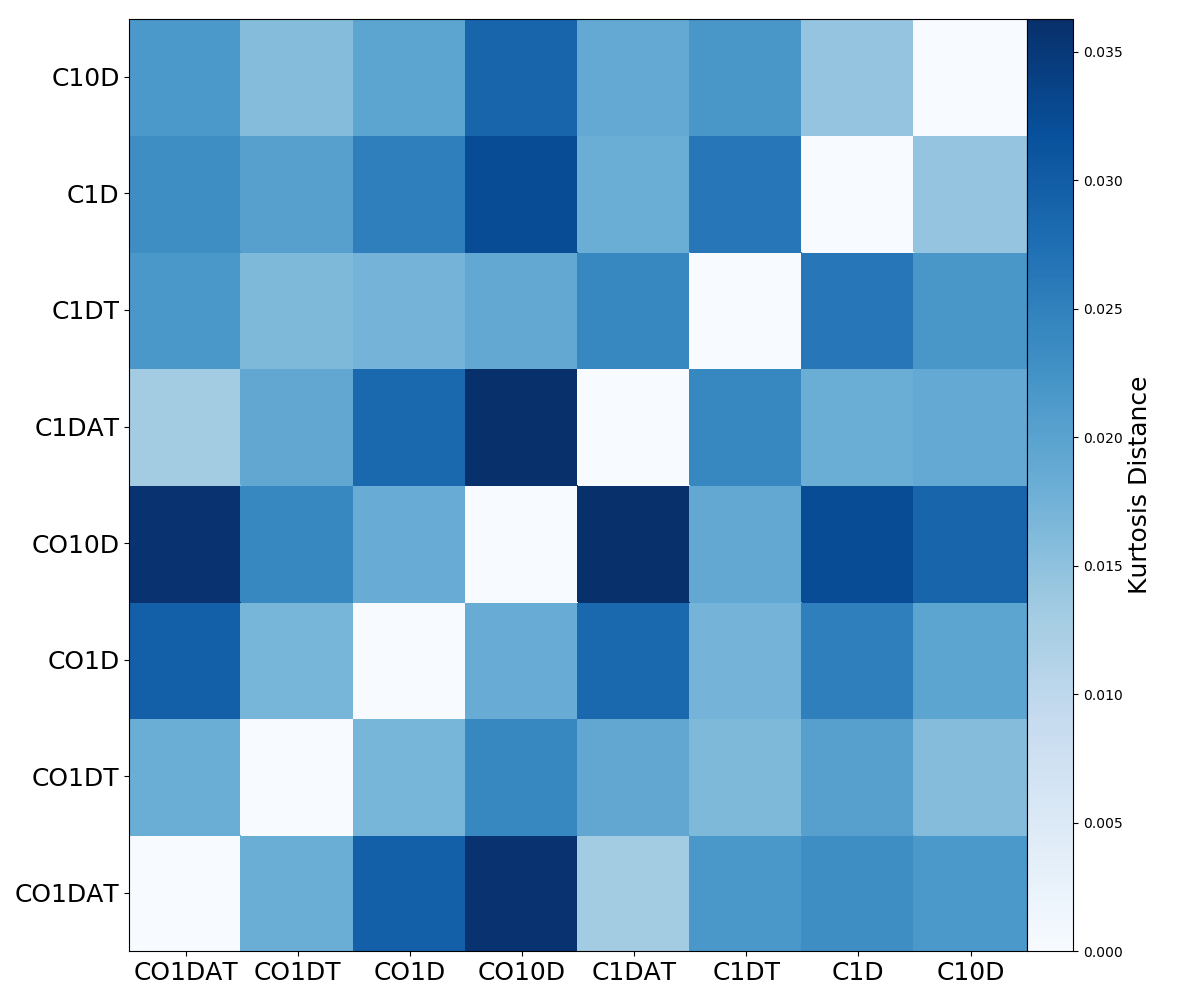} 
\includegraphics[width=0.95\columnwidth]{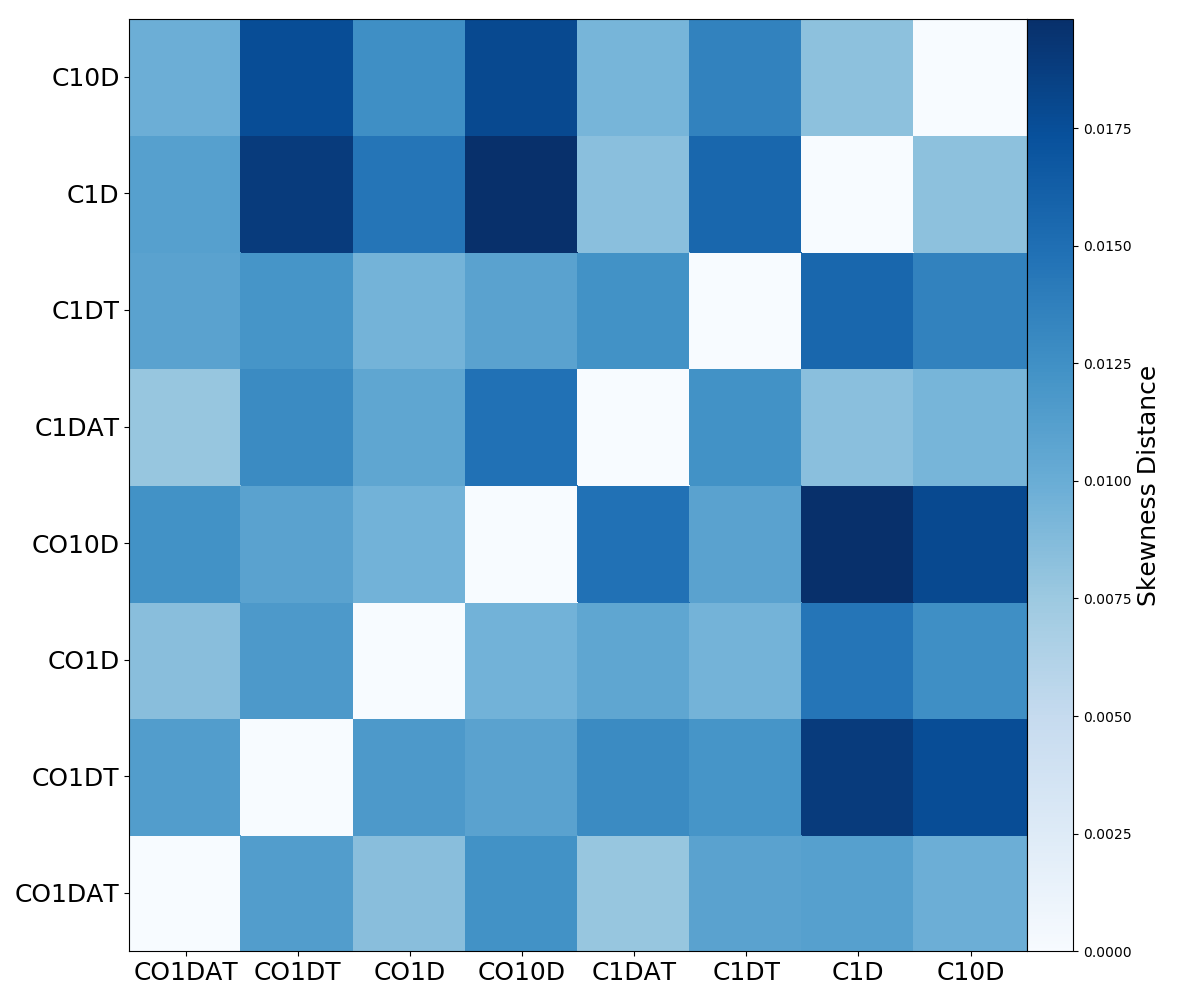} \hspace{0.1in} 
\includegraphics[width=0.95\columnwidth]{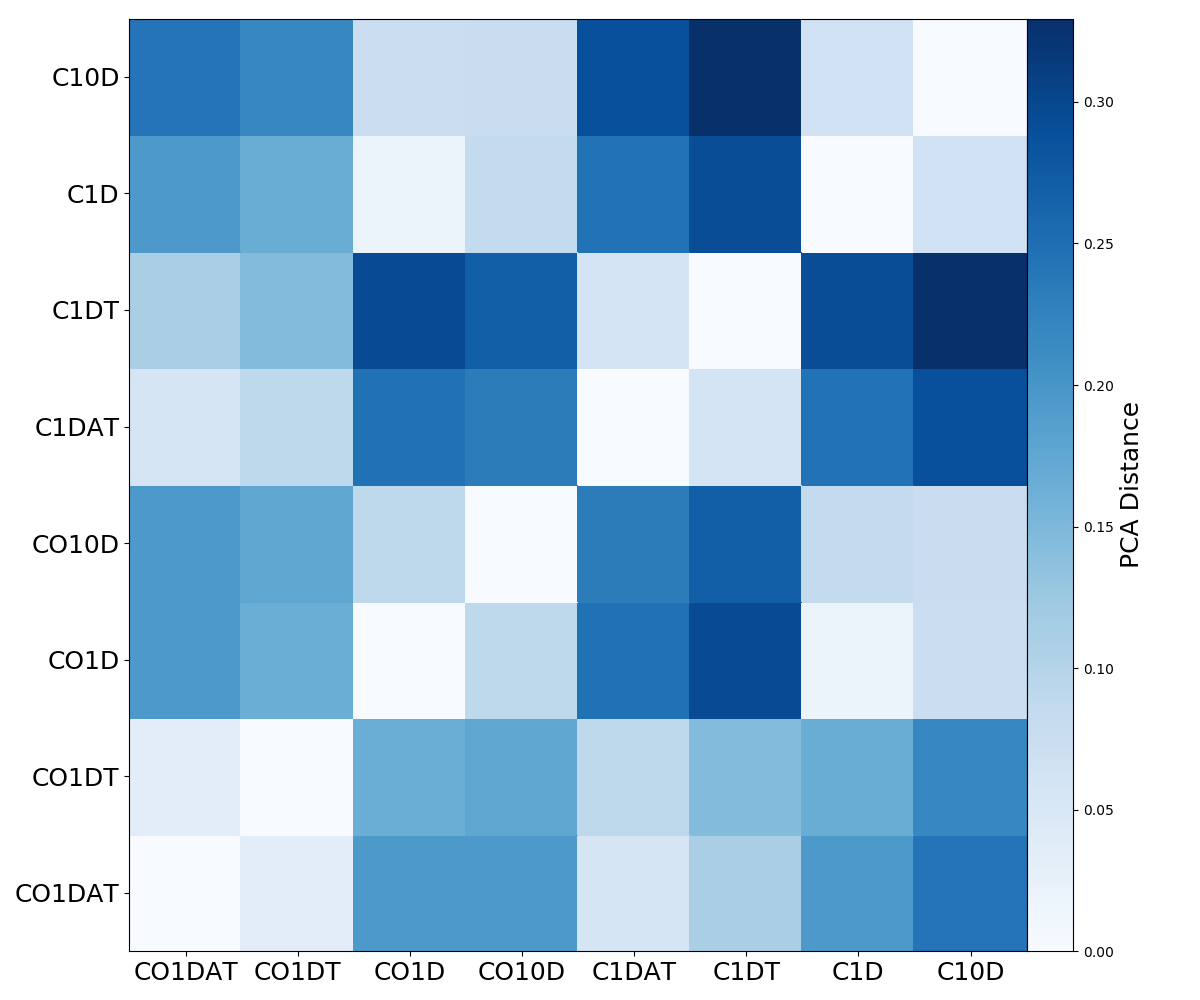}  
\includegraphics[width=1.0\columnwidth]{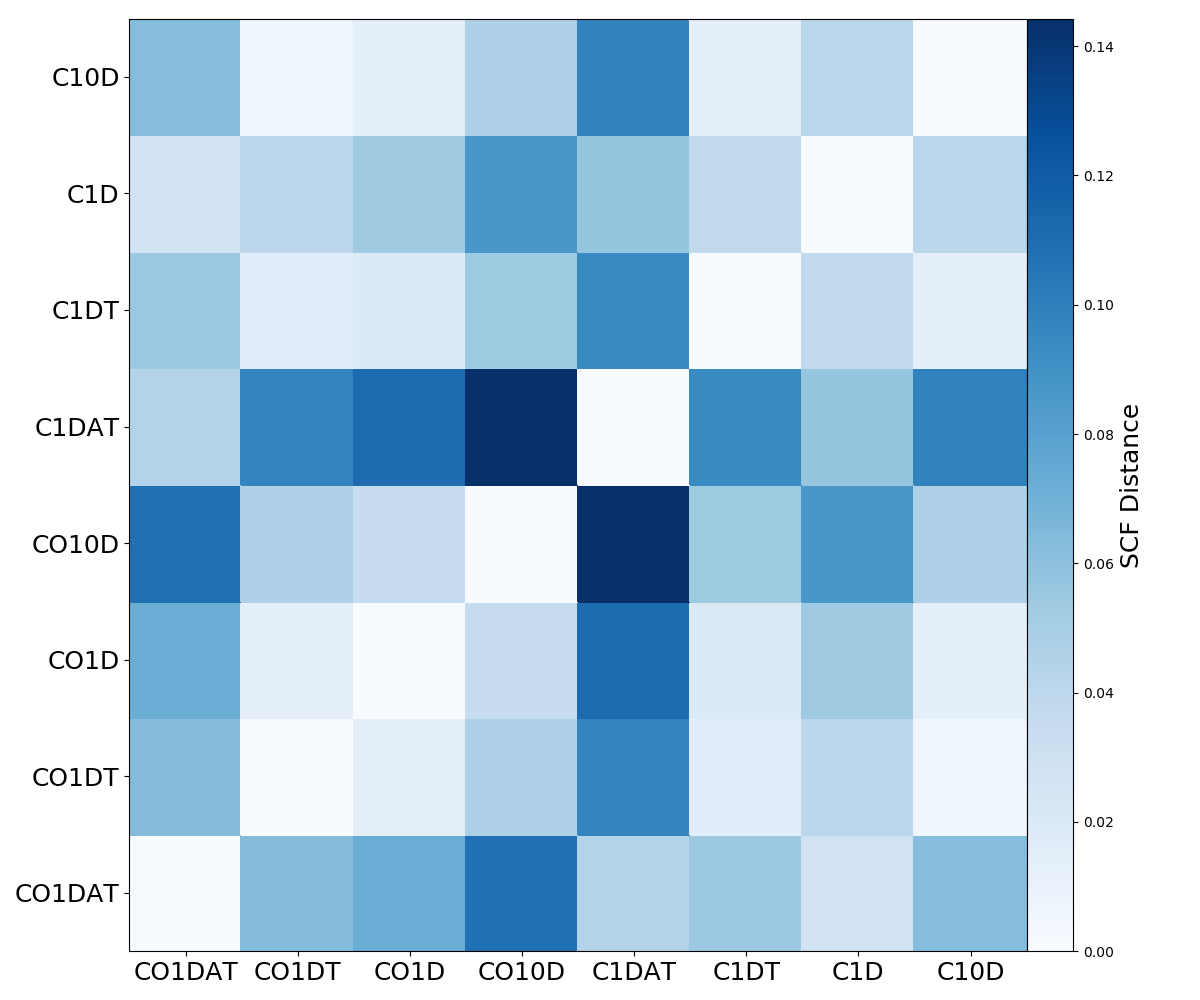} \hspace{0.1in} 
\caption{Intensity statistic color plots. Each statistic utilizes a different distance metric to quantify the difference between two models. The colored squares represent the distance between the models on the horizontal and vertical axes. \label{color_inten}}
\end{center}
\end{figure*}

Figure \ref{color_inten} displays the color plots for all intensity statistics. 
These statistics exhibit a wide range of responses to the model parameters. The PDF 
primarily responds to abundance variation. Models with fixed abundances have the largest lognormal widths, and models with varying abundances are consistently narrower (see Appendix \ref{appendix}). C models also show stronger responses to temperature variation and radiation than CO models, which exhibits weaker responses.
Overall, when we 
exclude the lower-intensity excess, significant distances are associated with  
changes in the PDF's Gaussian-like behaviors.

The kurtosis and skewness PDFs show sensitivity to all model parameters. The largest skewness distances arise from pairs with different gas tracers, but we find no meaningful trends in the skewness color plot. As shown in Figure \ref{highstats}, our models only slightly impact the amplitudes and widths of the skewness PDF. For the kurtosis PDF, the largest distances are associated with the interstellar field strength, and the CO models show increasing distances as we systematically introduce each model parameter. Given that no isolated strong parameter trends are present in the kurtosis color plot, we suggest that the kurtosis is also insensitive to our model parameters. Furthermore, interpreting both color plots is further complicated when considering that data sets with different spatial structures can produce similar kurtosis/skewness PDFs. However, the skewness and kurtosis may serve as a proxy for discontinuities in cloud structure.

The SCF distance metric, whose color plot resembles those of the higher-order moments, produces the largest distance from pairs just involving X1DAT. The corresponding spectra (see Figure \ref{scf_spectra}) have the largest SCF values over most lags, which increases the distances. The corresponding CO and CI models have small distances, which agrees with the expectation from \citet{gaches15}. 
Overall, these three statistics produce notably different responses for models with astrochemistry.

We find that the PCA color plot is checkered. The greatest distances occur between models with temperature variation and those without. Abundance variation, gas tracer  and field strength independently have little impact on the distance. This trend is largely driven by the decrease in relative variance encapsulated by the second eigenvalue, as shown in Figure \ref{pca_eigs}.

\subsection{Fourier Statistics}

\begin{figure*}[h!]	
\begin{center}				\includegraphics[width=0.95\columnwidth]{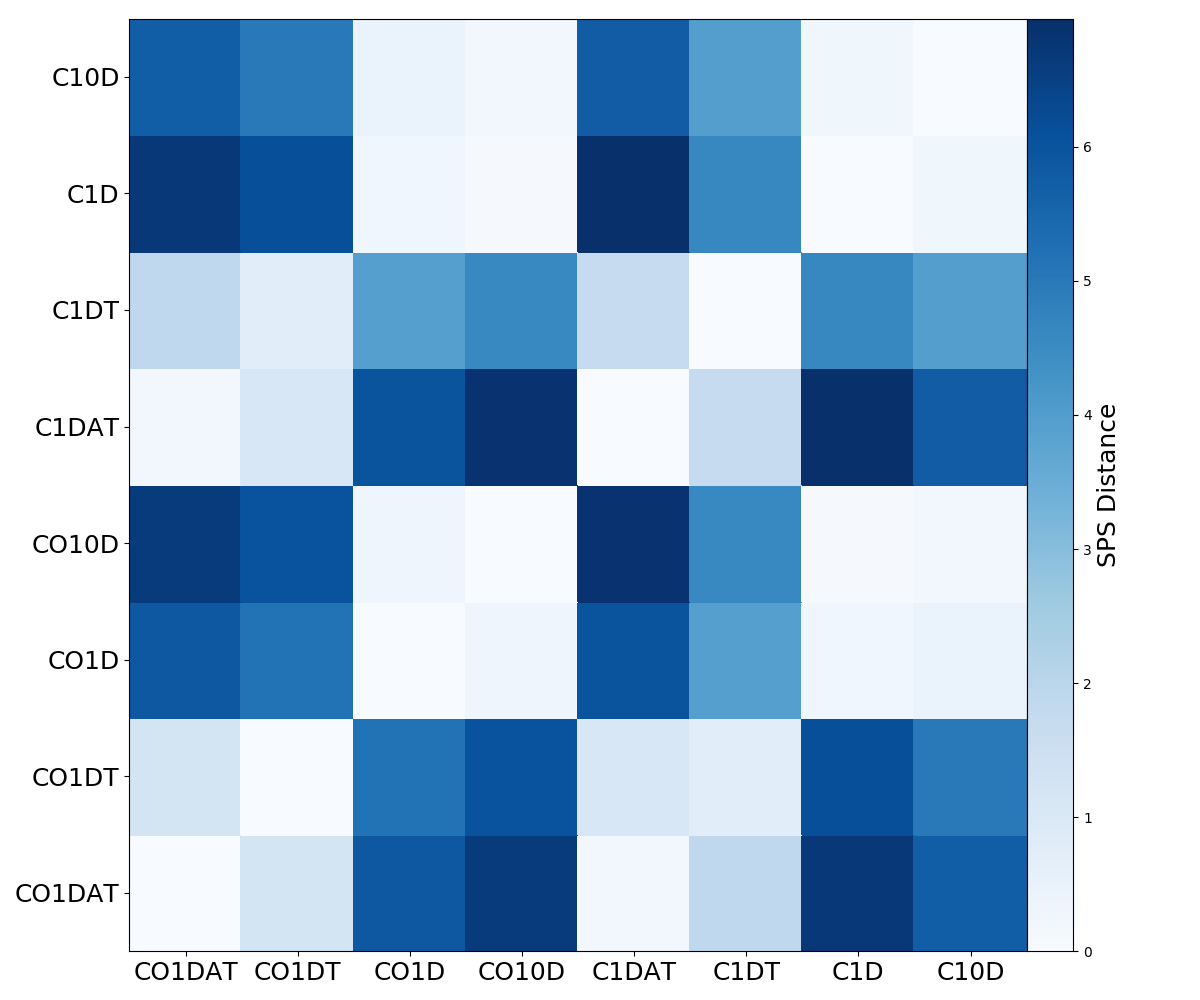} 
\includegraphics[width=0.95\columnwidth]{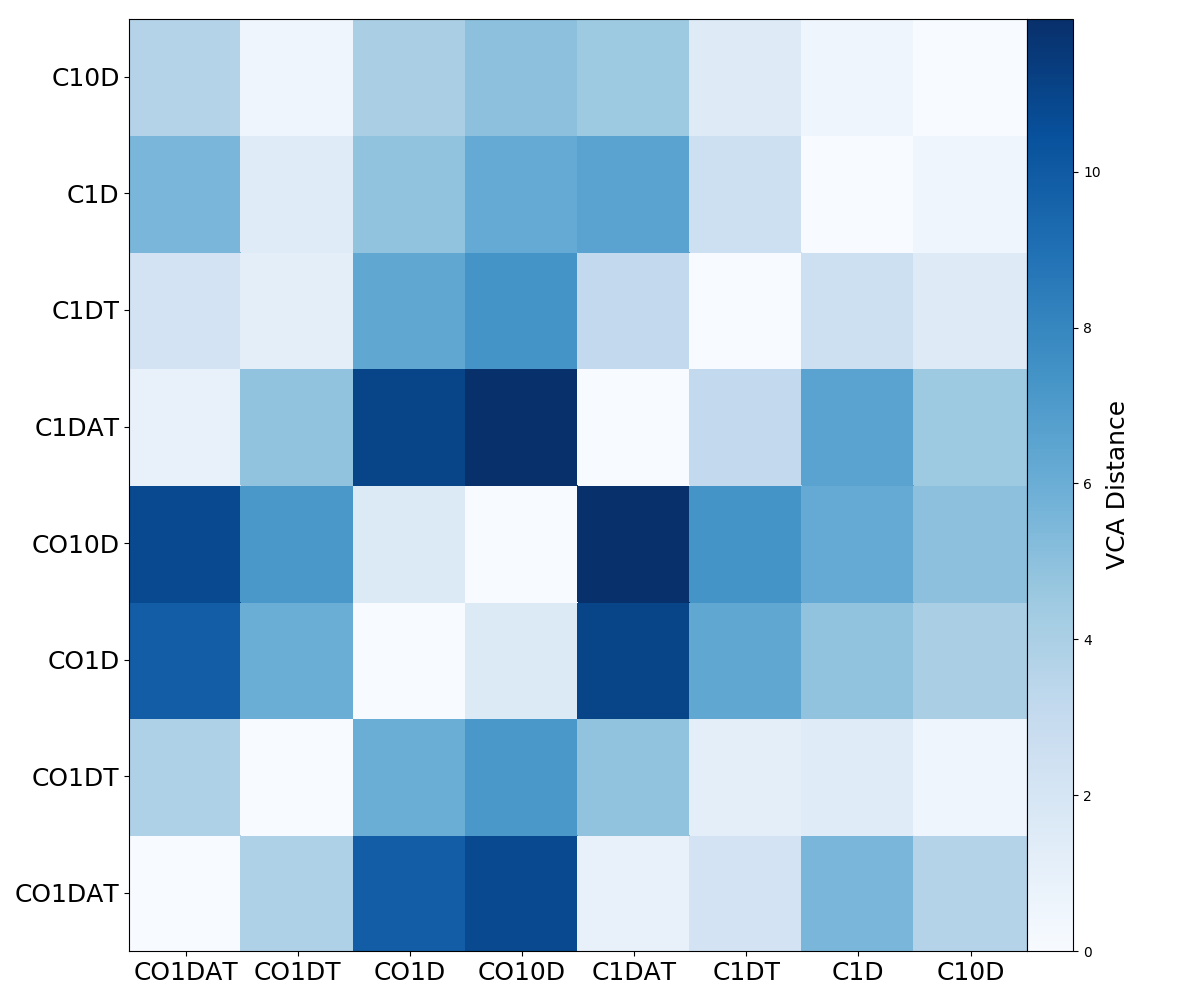} 
\includegraphics[width=0.95\columnwidth]{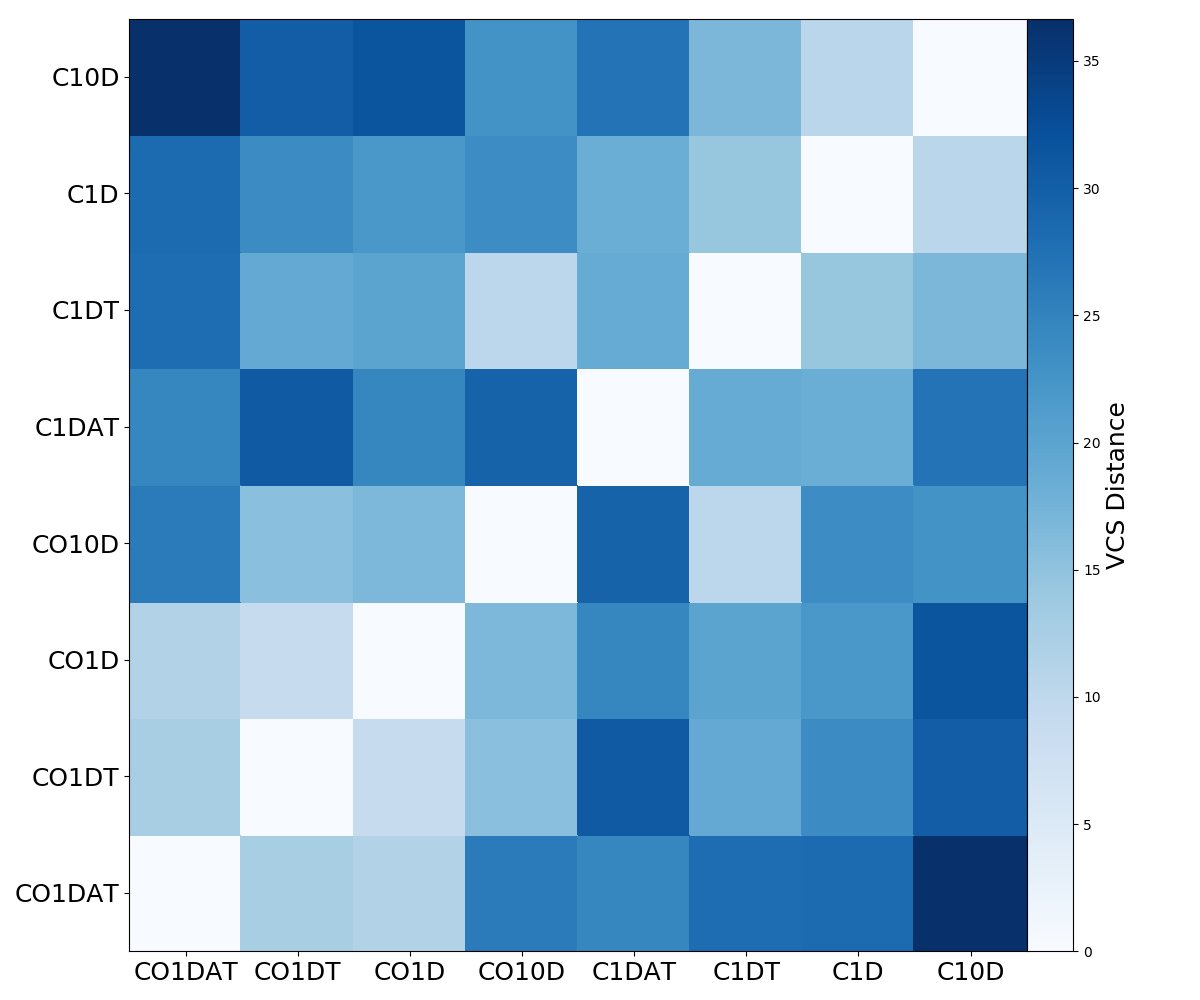} \hspace{0.1in} 
\includegraphics[width=0.96\columnwidth]{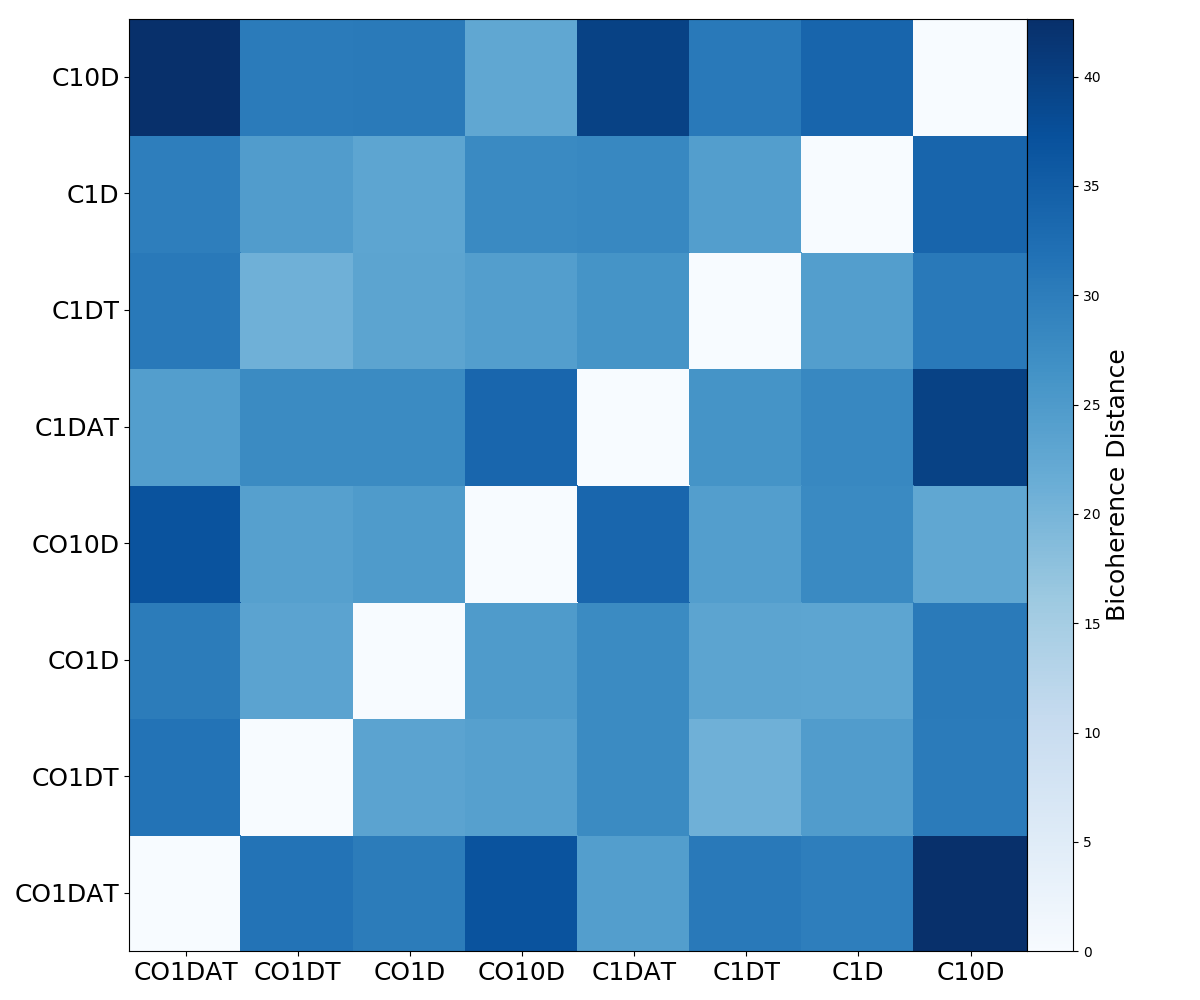}  
\includegraphics[width=0.95\columnwidth]{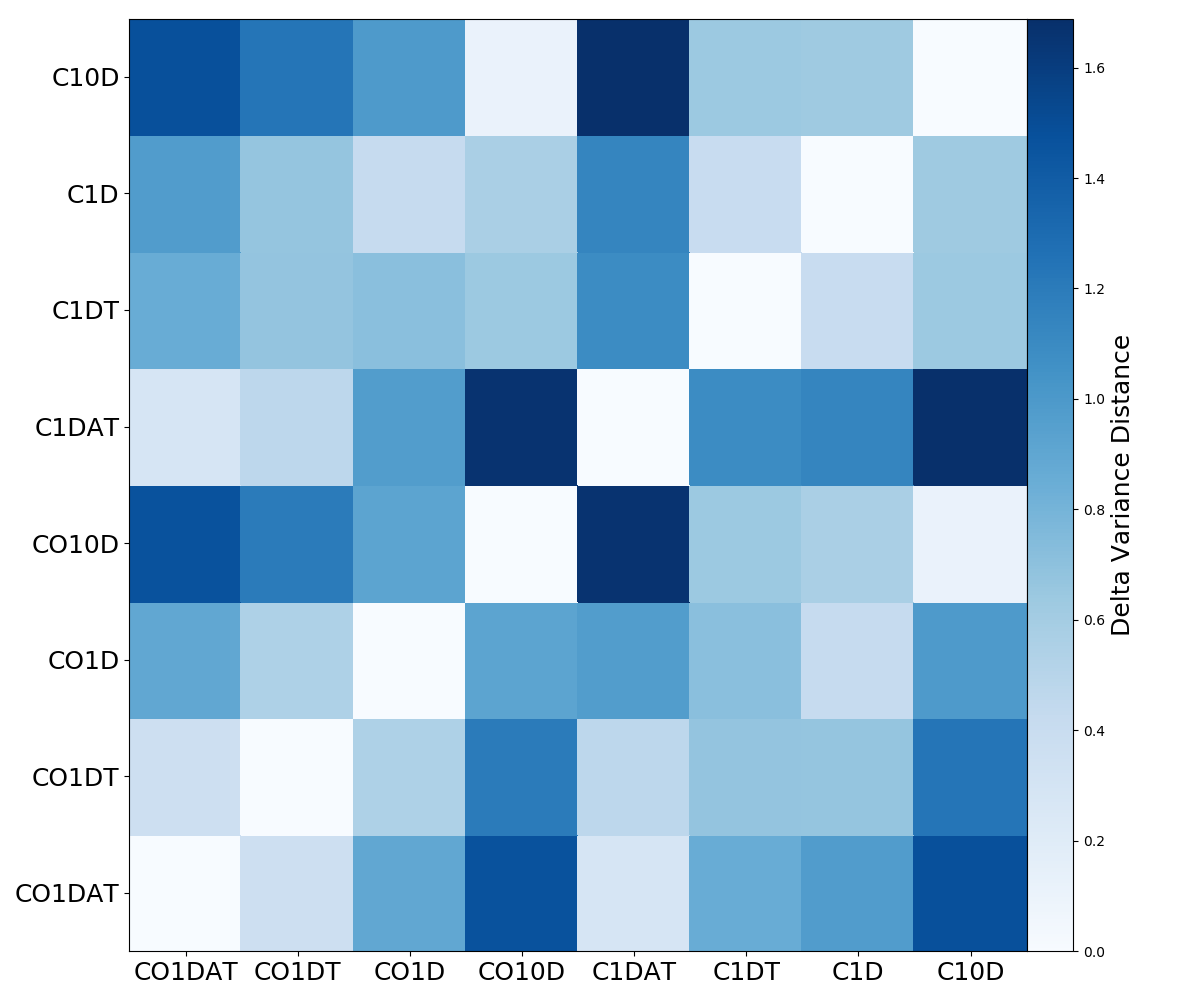} \hspace{0.1in} 
\includegraphics[width=0.95\columnwidth]{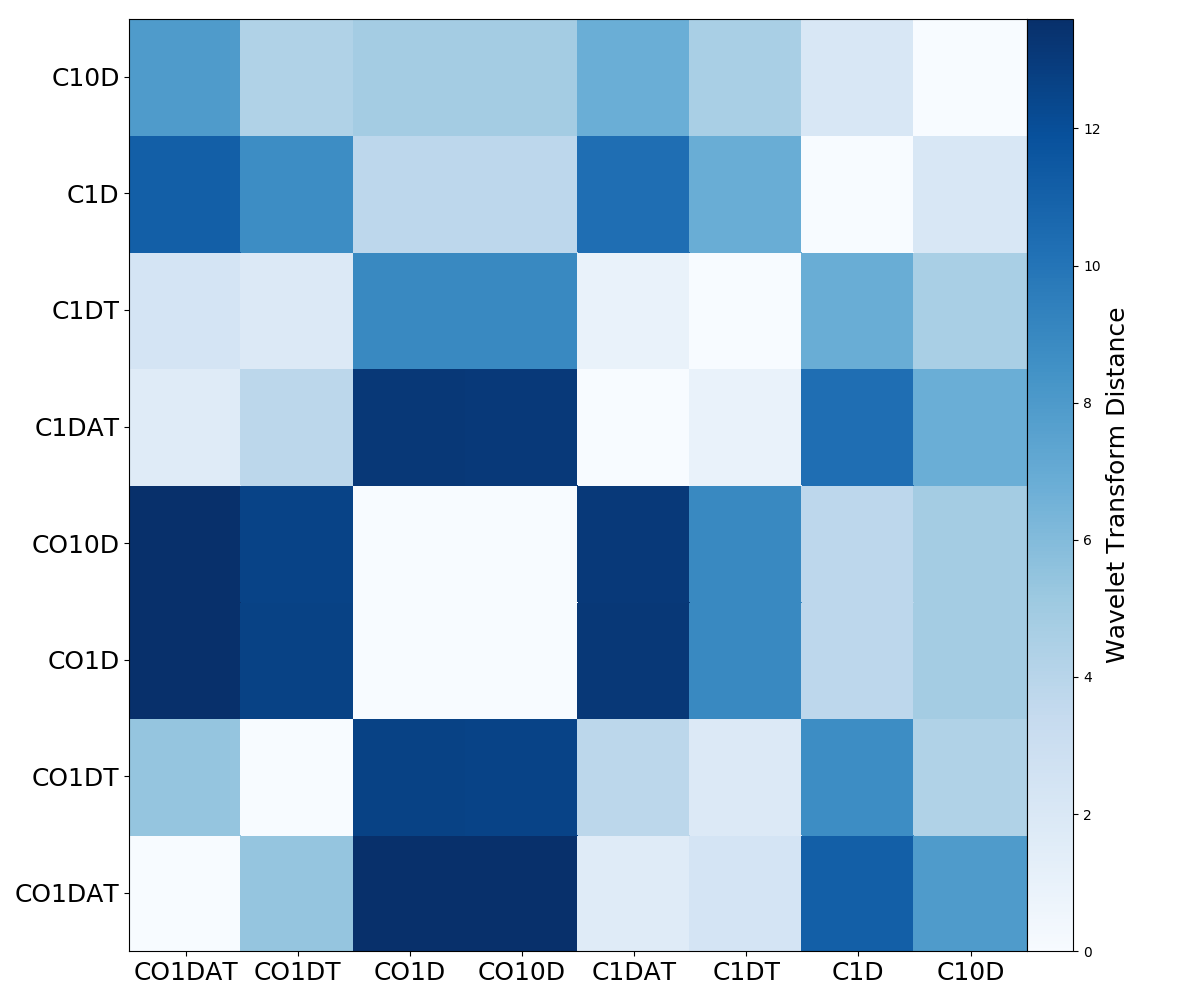}
\caption{Fourier statistic color plots.\label{color_fourier} }
\end{center}
\end{figure*}

Figure \ref{color_fourier} displays the color plots for all Fourier statistics. 
SPS and VCA, despite having similar definitions, show different sensitivities to our model parameters. We find that the SPS distance metric is most responsive to temperature variation, and slightly responsive to radiation field strength. For the VCA, CO models appear most sensitive to temperature variation, but this trend is unclear for C models. 

The bicoherence color plot shows sensitivity to all model parameters, and the largest distances arise from an amplified radiation field strength. Including chemistry and a stronger radiation field nearly removes all phase coherence, which makes our results progressively less correlated. 
Thus, when looking for specific trends in models or observations, we suggest that the bicoherence may be challenging to interpret.

The VCS color plot suggests a sensitivity to gas tracer, e.g., the largest distances appear in the top-left 4x4 block, which  
corresponds to comparisons between a CO model and a C model. Although the VCS distance metric is just based on the fitted slopes, it successfully accounts for differences between gas tracers. The metric responds to changes in radiation field strength for CO emission and to the inclusion of abundance and temperature variations for C emission. As shown in Figure \ref{vcs}, the sensitivities appear to be mostly associated with changes in the small-scale (larger $k_v$) slope.

The $\Delta$-variance shows responses to all model parameters. For both CO and C, the distances increase monotonically as we systematically introduce abundance variation, temperature variation and a stronger radiation field. The largest relative change in distance is associated with the stronger radiation field. We note no major differences associated with gas-tracer, which we previously in section \ref{delvar_sec}. Thus, we suggest that the $\Delta$-variance is most sensitive to the radiation field strength.

The wavelet transform power law behaves similarly to the SPS, but its color plot suggests a greater resemblance to the VCA. Namely, the CO and C models exhibit different behaviors. We find that CO models show a strong response to temperature variation and a very weak response to the radiation field strength. C models, however, exhibit a weaker response to temperature variation, and they appear more sensitive to the interstellar radiation field strength.

\subsection{Morphology Statistics}

\begin{figure}[h!]
\begin{center}
\vspace{0.3in} 
\includegraphics[width=0.95\columnwidth]{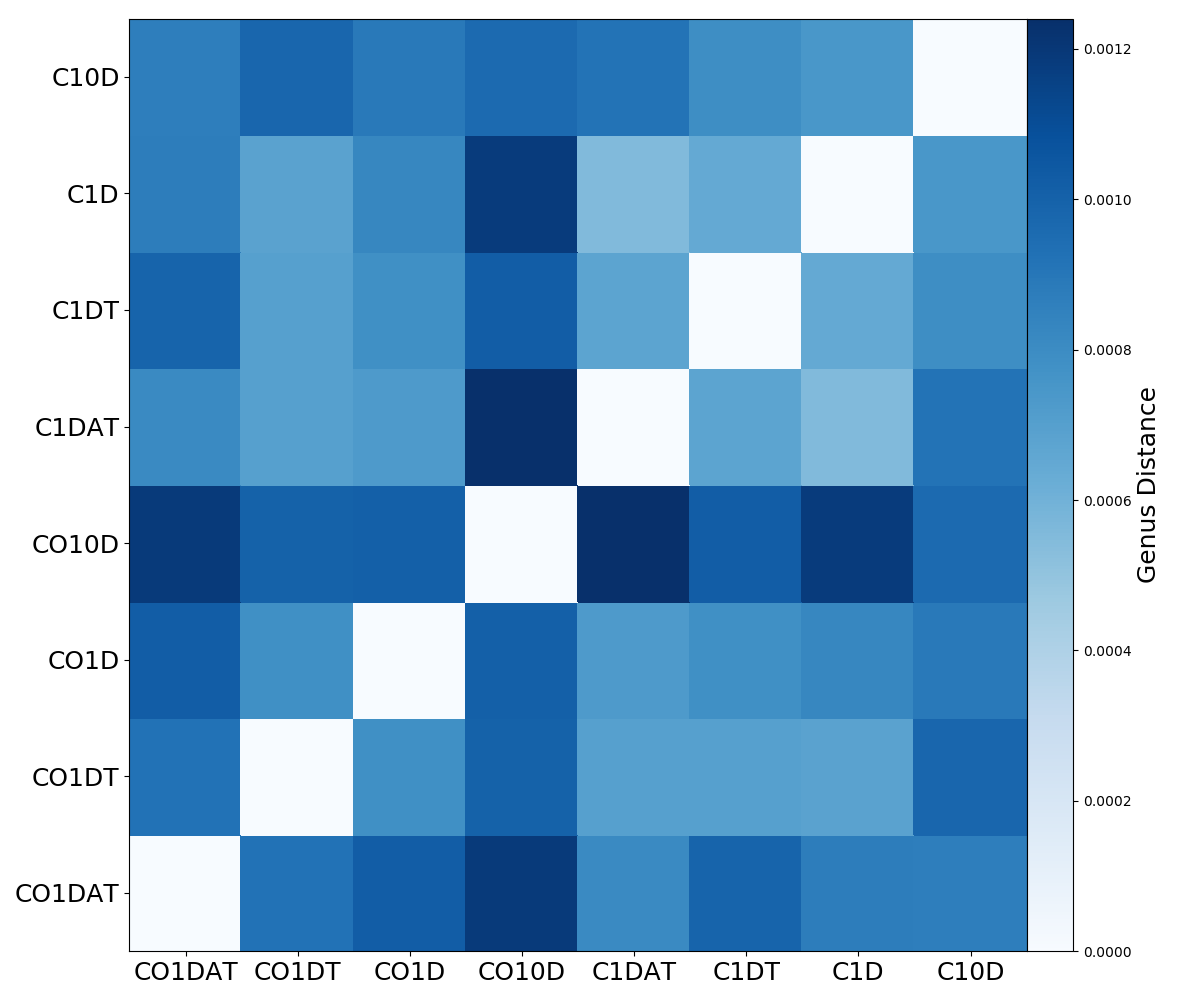}  
\vspace{0.1in} 
\includegraphics[width=0.95\columnwidth]{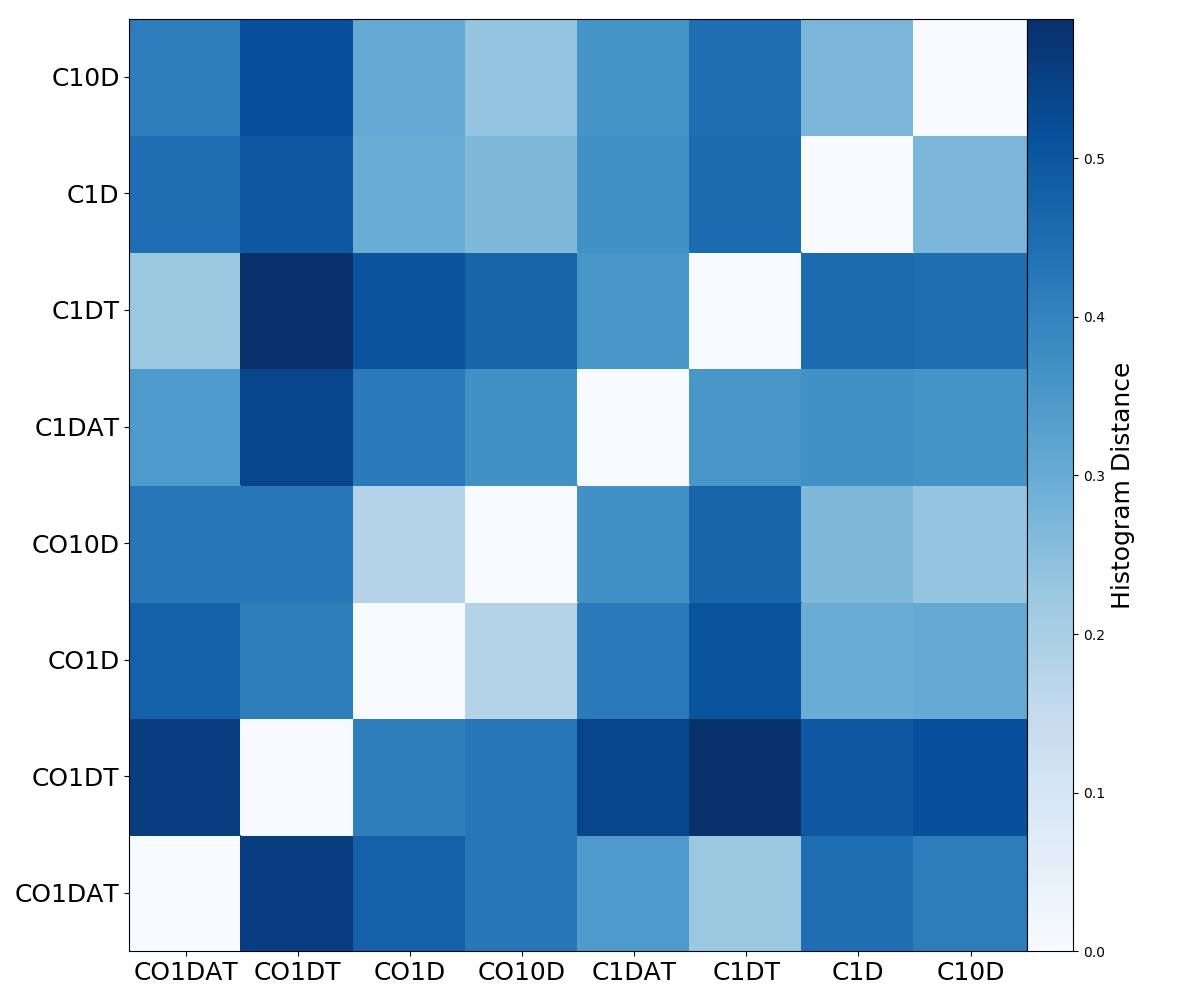} 
\vspace{0.1in} 
\includegraphics[width=0.95\columnwidth]{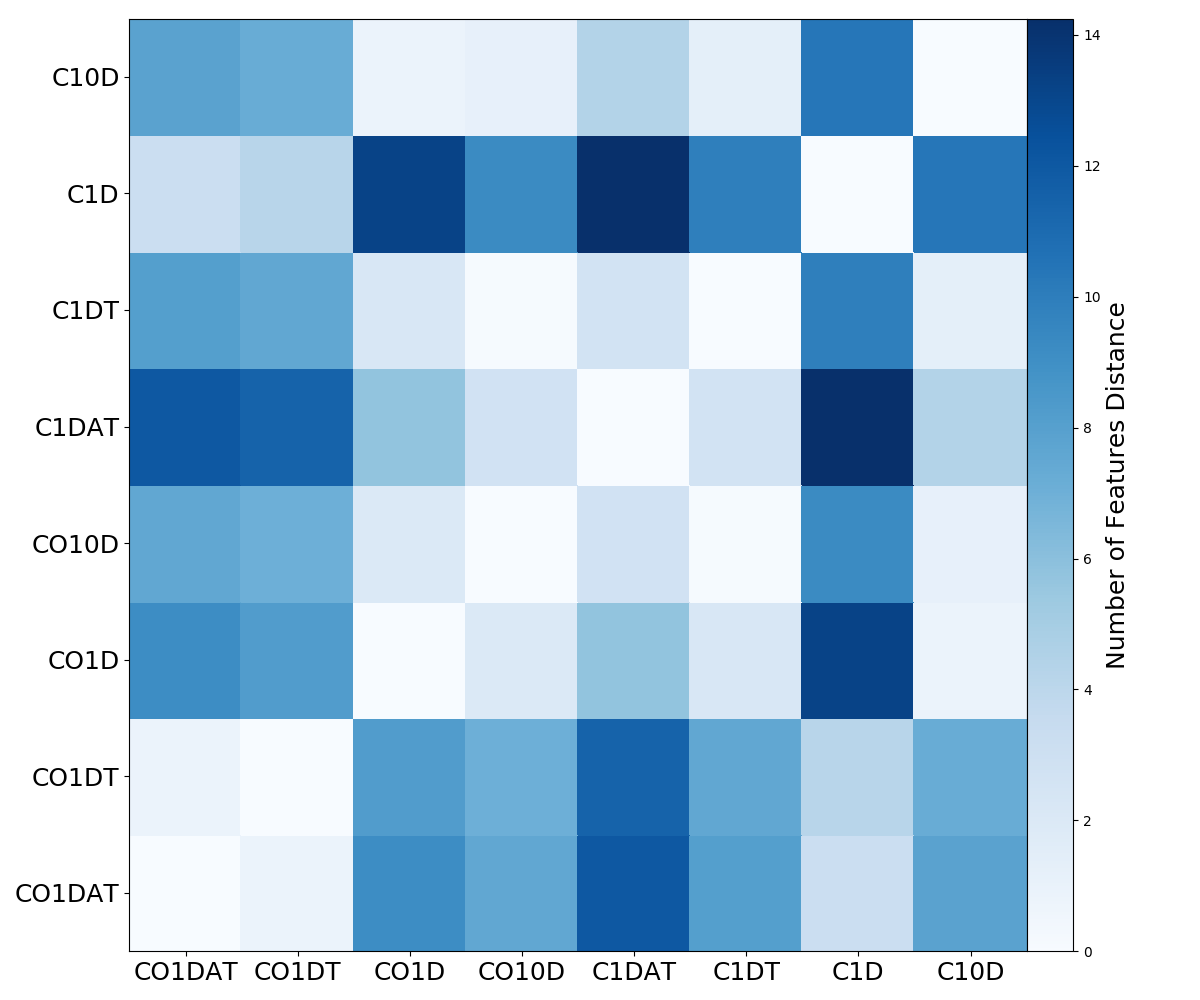} \vspace{-0.1in} 
\caption{Morphology statistic color plots.  \label{color_morph}
}
\end{center}
\end{figure}

Figure \ref{color_morph} shows the color plots for the morphology statistics. The genus statistic returns significant distances for nearly all model pairs.
Model CO1D has the largest distances; otherwise the distances between any two C models and C and CO pairs are similar.
B16 concluded the genus statistic did not have a strong sensitivity to wind feedback, and indeed, the distances are similar in the two analyses. As a result, we find this color plot to be challenging to interpret.

Like the genus statistic, the dendrogram histogram statistic is also challenging to interpret: it 
does not show clear trends with increasing complexity or stronger radiation field. The distances between CO pairs is similar to the distances between C and CO model pairs. Radiation field strength has the least impact on the histograms. B16 found the genus statistic is sensitive to the strength of wind activity, and the distances driven by feedback activity exceed the distances driven by radiation field and chemistry.

The dendrogram number of features distance metric only highlights differences in the size of the fit range, as expected. The largest distances arise from pairings that contain either CO1DAT, CO1DT, or C1D, which all have the shallowest slopes and the largest fit ranges. 
Thus, the number of features statistic is mainly sensitive to temperature variation, and in some instances, gas tracer. Including a stronger radiation field appears to mitigate the differences between CO and C models, i.e., C1D's number of features power law is quite different than that of CO1D, but the slopes of C10D and CO10D appear statistically indistinguishable. Currently, we do not know the exact radiation field strength in which the dendrograms continue to resemble to each other, so the distance metric's sensitivity to gas tracer may be challenging to interpret without additional study. 

\section{Discussion} \label{discuss}

\subsection{Impact of Astrochemistry in the Context of Prior Work}

To date, no paper has systematically investigated the impact of astrochemistry treatment on statistical metrics. Several studies have, however, analyzed 3D models of molecular clouds including abundances and temperatures calculated using chemical networks for individual statistics \citep[e.g.,][]{bertram14, bertram15b, gaches15}.

\citet{bertram14} used the PCA to study the impact of time-dependent chemistry,  radiative transfer and mean density on turbulent gas dynamics. Their MHD simulations did not include self-gravity, star formation or stellar feedback, and they 
focused on studying the distributions of H$_2$,  $^{12}$CO, and $^{13}$CO gas molecules. Using various initial gas densities, \citet{bertram14}  constructed position-position-velocity (PPV) cubes of the density distributions of all gas, of H$_2$, and of CO, as well as PPV cubes of the intensity distributions of $^{12}$CO and $^{13}$CO, both of which included the radiative transfer step. For each cube, they performed a PCA analysis to extract a pseudo-structure power-law that described the velocity fluctuations as a function of spatial scale. This methodology differed from B16 and K17, who only studied the PCA covariance matrix and eigenvalue set. 

 \citet{bertram14} measured distinct power-law slopes for the $^{12}$CO and $^{13}$CO intensity PPV cubes. In their analysis, PCA was not able to detect large-scale turbulent structure in the emission of the models with smaller initial densities ($n_{\rm H} = 30$ cm$^{-3}$).  For denser models ($n_{\rm H} = 300$ cm$^{-3}$),    
similar to the conditions in B16,
the PCA recorded structure on scales as large as half of the cube width. For the fiducial model, which has an intermediate density ($n_{\rm H} = 100$ cm$^{-3}$), a power-law fit could only be obtained for $^{12}$CO, which \cite{bertram14} claimed to be due to differences in optical depth. Although different chemical networks yielded distinct power-law slopes, \citet{bertram14} concluded that the differences were only marginal and that initial density was more influential than gas chemistry. However, this premise only holds when comparing $^{12}$CO and $^{13}$CO emission, or CO and H$_2$ density. Under our simulated conditions ($n_{\rm H} = 855$~cm $^{-3}$), we find that PCA is mainly sensitive to gas tracer choice 
and temperature variation, factors which also influence the line shape and optical depth. Thus, our results are consistent with \citet{bertram14}, given that our parameters encapsulate underlying changes in the line shape. 

In a second paper, \citet{bertram15b} carried out a similar analysis of the same simulations examining the $\Delta$-variance statistic. 
Instead of comparing different spectral shapes and horizontal offsets, which was done in B16, \cite{bertram15b} fitted a region of each spectra to a power law and used the power-law slopes to derive line~width-size relations, quantifying the relationship between structure size and turbulent velocities. 
\citet{bertram15b} found that the $\Delta$-variance statistic, although strongly affected by optical depth, serves as a useful tool for extracting information about the molecular gas distribution. For their dense-gas models, they reported near-identical linewidth-size relations for the $^{12}$CO and $^{13}$CO emission PPV cubes, which suggests an insensitivity towards CO chemistry. They also argued that, below some critical density, CO loses effectiveness as a molecular gas tracer. 
Our analysis confirms that $\Delta$-variance is insensitive to the gas tracer, in the sense that a given $^{12}$CO model has a similar distance when compared to the another $^{12}$CO or C model, i.e., the CO1DAT and CO10D pair and the CO1DAT and C10D pair have nearly identical distances. Nonetheless, we find that the $\Delta$-variance registers clear changes to both the external radiation field and underlying variation in abundance and temperature. 
If CO and C trace some of the same gas (see \ref{codiss}), then the line shapes will appear similar for the same models even if the brightness differs. 

Finally, \citet{gaches15} deduced that the SCF is sensitive to gas tracer and  critical density. Using hydrodynamic simulations post-processed with {\sc 3d-pdr}, they generated synthetic PPV cubes of 16 common molecular cloud tracers, including CO and C, and calculated the corresponding SCF spectral slopes. 
\citet{gaches15} were able to classify tracers as diffuse, intermediate, or dense, based on the SCF slope, where denser tracers had steeper SCF slopes. 
In this framework, CO and C are both diffuse tracers with statistically similar SCF slopes. Likewise, our SCF formalism produces similar distances between CO and C pairs. Underlying this similarity, we find the SCF is independently sensitive to the strength of the radiation field and inclusion of chemistry; the largest distance is between CO10D and C1AT. 

\subsection{CO and C Emission Similarity} \label{codiss}

A variety of prior work has noted the qualitative correspondence between CO and C emission data \citep{papadopoulos04,Offner14b,glover15}. This suggests the two species have significant overlapping spatial distributions and trace similar subsets of H$_2$ number density. However, the similarity is  puzzling at face value because of their different critical densities and chemistries, especially given that CI emission bears a greater resemblance to $^{13}$CO(1-0), which tracers higher densities than $^{12}$CO(1-0) \citep[e.g.,][]{oka01}. 
The most plausible explanation is that complex cloud morphology facilitates more effective UV irradiation of the cloud interior and produces more overlap than expected compared to one-dimensional PDR models \citep{papadopoulos04,Offner14b,glover15}.

Turbulent statistics provide one way to probe and quantify the degree of similarity in different regimes.
For example, \citet{gaches15} showed that the SCF slope of CO and C emission cubes exhibits quantitative agreement, where the difference in slope between the two tracers is within the range of variation for different viewing angles for each individually. Of the statistics examined in this work, the PDF, skewness, kurtosis, PCA, 
SCF, bicoherence, 
SPS, $\Delta$-variance and wavelet transform show the most quantitative similarity between the CO and C emission pairs.  This manifests as similar trends in distances between CO pairs and CO and C pairs, i.e., the bottom left and bottom right $4\times 4$ matrices in the color plots appear similar. Of particular interest, VCS and PCA 
show this trend while {\it also} reporting notably larger distances between CO-C pairs than the complementary CO-CO model pairs: CO-C pairs are similar to one another but more dissimilar than the same CO-CO pair.
Statistics that do not show similarity between CO and C emission tend to exhibit non-monotonic responses to the other parameters as well. These include the 
VCA, genus and both dendrogram statistics. 

\subsection{Implications for Identifying Feedback and Physical Variation between Clouds} 

Of the intensity statistics, B16 identified the PCA and SCF as the first- and second-most promising candidates for identifying stellar feedback. However, those  models do not incorporate gas chemistry. 
Comparing with B16, we find the maximum PCA distances between wind and non-wind runs are similar to those between chemistry and non-chemistry models. However, B16 found that stellar feedback creates strong covariance for velocities $|v| \sim$ 2-3 km~s$^{-1}$, i.e., a unique signature associated with the characteristic velocity of the winds. Thus, we expect PCA to continue to be a robust indicator of feedback. 

The maximum SCF distances due to the inclusion of chemistry and feedback are also similar. However, winds steepen the SCF slope while chemistry flattens it.
For kurtosis and skewness, 
the distances between simulations with and without feedback are a factor of $\sim$ 1 and 2,
respectively, 
smaller than distances due to chemistry. 
The PDF distances in this paper are significantly larger than those in B16. However,  here we use a new definition for the PDF distance metric. Considering that both feedback (see B16) and chemistry yield narrower PDF widths, we expect these two parameters to have comparable effects on the PDFs. 
These findings
suggest that chemical modeling may be crucial for constraining the impact of feedback in comparisons between simulations and observations for the SCF and PDF. 

B16 found both VCS and $\Delta$-variance statistics also exhibit moderate sensitivity to feedback. However, we note the maximum distances due to feedback are comparable to those between X1DAT and X10D pairs here. This suggests that uncertainties in the local radiation environment may be difficult to separate from differences caused by feedback. B16 concluded that the Fourier statistics failed to produce unique feedback signatures, although several distance metrics registered sensitivity to feedback. Since chemistry does impact the distances, including it may produce clearer statistical responses to feedback and overall larger distances.

Although B16 classified both dendrogram statistics as effective indicators of feedback, here neither of the statistics respond monotonically to changes. Thus, the feedback diagnostics for the histograms identified in B16 may change 
when chemistry is included.

\subsection{Caveats} 

While this work extends \citet{yeremi14}, B16 and K17, we note a variety of caveats to drawing generalized conclusions concerning the merits of the different statistical approaches.

Our astrochemical network is relatively advanced and includes thousands of reactions; however, it is fundamentally a network for treating photodissociation regions. It does not treat the freeze-out of CO onto dust grains, which occurs in cold, dense gas, or shock chemistry. These effects are important for studies of dense cores and star-forming clouds with vigorous protostellar feedback. However, in the present study high-density gas composes $\lesssim 1$\% of the volume and the freeze-out of CO occurs after the lines are already optically thick. Consequently, we expect our results to be reasonably robust and applicable to typical Milky Way clouds. Future work is needed to study the interplay of feedback and chemistry.  

Second, our current study includes only a single turbulent realization. This is by construction since it allows us to isolate the impact of abundance, temperature and external field on each statistic. Consequently, our distance metrics likely represent lower limits, e.g., CO1DAT likely exhibits a larger distance from CO10D with a different underlying turbulence pattern. However, we expect the reported trends to hold for different views and turbulent seeds based on the detailed investigation of K17. K17 showed all the statistics we include in our analysis are reasonably insensitive to statistical fluctuations including added random noise. Of these, we note K17 found VCS, the bispectrum and $\Delta-$variance, when applied to the integrated intensity, are less robust (see Table 3 in K17 for the p-values of each), a detail worth bearing in mind for quantitative comparisons.

Finally, we note that several statistics, such as PCA and SCF, have different definitions in the literature. Our conclusions apply only to the particular formulation implemented in {\sc turbustat}. 

\section{Conclusion}\label{conclusions}

We investigated the impact of astrochemistry on 15 common turbulence statistics: intensity PDF, skewness, kurtosis, power spectrum, PCA, SCF, 
bispectrum, VCA, VCS, $\Delta$-variance, wavelet transform, genus, number of dendrogram features, and the histogram of dendrogram feature intensities. We analyzed the hydrodynamic simulations performed by \cite{Offner13} to build a 
suite of synthetic observations of turbulent, star-forming molecular clouds. In the models, we evaluated three chemistry parameters: the degree of chemical complexity, the type of gas tracer, and the strength of the interstellar radiation field. We used {\sc 3d-pdr} to model chemical reactions, set the radiation field strength, and directly compute local gas abundances and temperatures. We computed the $^{12}$CO(1-0) and CI($^{3}$P$_{1}$-$^{3}$P$_{0}$) emission in radiative transfer processing, and we converted the emission into synthetic observations mimicking observations of the nearby Perseus molecular cloud. 

For each statistic, we compared all models and identified qualitative trends. 
We then used distance metrics to quantify the statistical responses and efficiently compare differences between various models. From these, we identified and determined a variety of sensitivities towards the chemistry parameters. We present the following conclusions:
\begin{enumerate}
\item At low intensities, the
PDF is highly sensitive to the degree of chemical complexity. Including chemistry creates a significant low-intensity spike
that overshadows the impact of other chemistry parameters. 
When we include a low-intensity cutoff in the distance calculations, we find clear sensitivities to abundance variation, radiation field strength, and gas tracer. 
\item The skewness and kurtosis PDFs are weakly sensitive to all model parameters, and the spatial distributions of skewness and kurtosis appear well-correlated with cloud discontinuities.
\item PCA is sensitive to chemical complexity in a binary way. 
Specifically, it is sensitive to temperature variation, but not variation in the local abundance. 
\item The SCF is sensitive to radiation field strength and abundance variations, both of which cause the SCF slope to flatten.
\item The SPS and VCA statistics yield similar power-law behaviors but different distance metric responses. SPS is highly responsive to temperature variation, but the VCA only responds to temperature variation for CO models. Both statistics also show weaker responses to radiation field strength. 
\item The VCS responds to all chemistry parameters but is primarily sensitive to the gas tracer. CO and C emission produce different VCS spectra, particularly at larger $k_v$ and at the break points.
\item The bicoherence displays equal sensitivity towards chemistry and radiation field strength. It is insensitive to differences in CO and C emission; in fact, CO and C outputs are nearly indistinguishable. 
\item The $\Delta$-variance responds most strongly to the radiation field strength. It shows weak but noticeable responses to abundance variation and temperature variation. The $\Delta$-variance spectra of CO models tend to peak at larger spatial scales than those of C models; however, differences between CO and C are not strong enough to produce a signal in the $\Delta$-variance distance metric.
\item The wavelet transform is responsive to temperature variation at small scales. At larger scales, the wavelet transform is sensitive to abundance variation and the radiation field strength. The main difference between the CO and C emission is the peak intensity; however, by design the wavelet transform distance metric does not depend on the signal strength.
\item The genus statistic shows no special sensitivity to chemistry parameters and produces significant differences for all model pairs.
\item The dendrogram statistics show some 
sensitivity to chemical complexity but the distances do not respond monotonically to gas tracer, increasing complexity or radiation field. 
\end{enumerate}

In summary, astrochemistry impacts the
PDF, PCA, 
SPS, VCA, VCS, bicoherence, $\Delta$-variance, wavelet transform, and dendrograms. 
In order to accurately interpret 
the statistical signatures of molecular cloud properties and to directly compare simulations and observations, we recommend including chemical modeling. Our results suggest accounting for the local radiation environment is especially crucial. 
The results of 
prior work that excluded chemistry may be affected. 
Our analysis, together with B16 and K17,  
underscore the importance of the input assumptions to the responses of turbulence statistics.

\acknowledgments

We acknowledge the helpful discussions of Ron Snell, Brandt Gaches and Mark Heyer, all of which improved our work. We also acknowledge helpful comments from two anonymous referees.
RB also thanks the University of Massachusetts Commonwealth Honors College for support in producing this work and acknowledges the support of Thomas Moser of Tall Corn Consulting, for readily assisting with software installations and providing necessary computational resources. SO acknowledges support from NSF AAG grant AST-1510021. EWK and EWR are supported by NSERC of Canada through graduate fellowships and a Discovery Grant.

\software{ORION \citep{Li12}, 3D-PDR \citep{Bisbas12}, RADMC-3D \citep{dullemond12}, TURBUSTAT \citep{Koch17}, Astropy \citep{Astropy}.  }

\bibliography{references,outflowbib}
\bibliographystyle{apj}

\appendix

\section{Statistical Fits} \label{appendix}

In Tables \ref{data1a} and \ref{data2}, we show all fit results for our statistical analyses.

\begin{deluxetable}{l | ccccc}
\tablecaption{Model fits\tablenotemark{a} \label{data1a}}
\tablehead{ \colhead{Model} \vline &  
   \colhead{PDF lognormal width} &
   \colhead{SCF power-law slope} &
   \colhead{SPS power-law slope} &
   \colhead{VCA power-law slope} &
   \colhead{Wavelet Transform power-law slope}
   }
\startdata
CO1DAT  & $0.576 \pm 0.002$ & $-0.132 \pm 0.006$ &  $-3.21 \pm 0.05$  & $-2.97 \pm 0.03$ & \textbf{$0.61\pm 0.02$} \\
CO1DT   & $0.458 \pm 0.002$ & $-0.134 \pm 0.004$ &  $-3.12 \pm 0.04$ & $-2.78 \pm 0.03$ & \textbf{$0.58 \pm 0.01$} \\
CO1D    & $0.430 \pm 0.001$ & $-0.146 \pm 0.004$ &  $-2.82 \pm 0.04$ & $-2.54 \pm 0.02$ & \textbf{$0.41 \pm 0.01$} \\
CO10D   & $0.439 \pm 0.002$ & $-0.149 \pm 0.009$ &  $-2.80 \pm 0.03$ & $-2.48 \pm 0.03$ & \textbf{$0.43 \pm 0.01$} \\
C1DAT   & $0.605 \pm 0.002$ & $-0.103 \pm 0.006$ &  $-3.19 \pm 0.05$ & $-3.01 \pm 0.03$ & \textbf{$0.60 \pm 0.01$}\\ 
C1DT    & $0.409 \pm 0.002$ & $-0.148 \pm 0.004$ &  $-3.07 \pm 0.05$ & $-2.85 \pm 0.04$ & \textbf{$0.56 \pm 0.02$} \\
C1D     & $0.264 \pm 0.001$ & $-0.122 \pm 0.003$ &  $-2.81 \pm 0.03$ & $-2.72 \pm 0.03$ & \textbf{$0.50 \pm 0.01$}\\
C10D    & $0.336 \pm 0.002$ & $-0.147 \pm 0.004$ &  $-2.79 \pm 0.05$ & $-2.75 \pm 0.05$ & \textbf{$0.50 \pm 0.01$} \\
\enddata
\tablenotetext{a}{Fit estimates and uncertainties for the PDF, SCF, SPS, VCA, and wavelet transform. The PDF lognormal width is calculated with a maximum likelihood estimator. All power-law slopes are obtained using an ordinary least squares method.} 
\end{deluxetable}

\begin{deluxetable}{l | cccc}
\tablecaption{Model fits, continued\tablenotemark{a} \label{data2}}
\tablehead{ \colhead{Model} \vline &  
   \colhead{VCS power-law slope at lower k$_v$}\tablenotemark{b} &
   \colhead{VCS power-law slope at higher k$_v$} & 
   \colhead{VCS break point (km/s)$^{-1}$} &
   \colhead{Dendrogram feature number power-law slope}
   }
\startdata
CO1DAT    & $-1.92 \pm 0.02$ & $-2.97 \pm 0.05$ & $-0.06 \pm 0.01$ & $-1.63 \pm 0.03$ \\
CO1DT     & $-1.62 \pm .02$  & $-3.17 \pm 0.03$ & $-0.12 \pm 0.01$ & $-1.66 \pm 0.06$ \\
CO1D      & $-1.78 \pm 0.02$ & $-3.39 \pm 0.04$ & $-0.03 \pm 0.01$ & $-2.04 \pm 0.04$ \\
CO10D     & $-1.36 \pm 0.03$ & $-3.65 \pm 0.05$ & $-0.13 \pm 0.01$ & $-2.19 \pm 0.08$ \\
C1DAT     & $-2.21 \pm 0.03$ & $-4.38 \pm 0.07$ & $0.03 \pm 0.02$ & $-2.46 \pm 0.06$ \\ 
C1DT      & $-1.48 \pm 0.03$ & $-4.19 \pm 0.05$ & $-0.10 \pm 0.02$ & $-2.21 \pm 0.07$ \\
C1D       & $-1.69 \pm 0.04$ & $-5.18 \pm 0.08$ & $0.02 \pm 0.02$ & $-1.53 \pm 0.01$ \\
C10D      & $-1.43 \pm 0.05$ & $-6.0 \pm 0.1 $ & $-0.02 \pm 0.03$ & $-2.09 \pm 0.05$ \\
\enddata
\tablenotetext{a}{Fitted power-law slopes and uncertainties for the VCS and Dendrogram feature number, which we calculate with ordinary least squares methods.} 
\tablenotetext{b}{Since the VCS fit is a segmented power-law, we report the fit information in three separate columns: the fit at lower velocity-frequencies, the fit at higher velocity-frequencies, and the break point between the two power laws.}
\end{deluxetable}

\end{document}